\documentclass[10pt,a4paper]{article}
\pdfoutput=1
\usepackage[english]{babel}
\usepackage[latin1]{inputenc}
\usepackage{amsfonts,amsbsy,bm,euscript,mathrsfs}
\usepackage{amssymb,stmaryrd,faktor}
\usepackage[tbtags]{amsmath}
\usepackage{graphicx}
\usepackage[title,titletoc]{appendix}
\usepackage[bookmarks=true,colorlinks=true,linkcolor=blue,citecolor=blue,urlcolor=blue,bookmarksnumbered]{hyperref}

\numberwithin{equation}{section}

\makeatletter
\renewcommand\section{\@startsection {section}{1}{\z@}
{-3.5ex \@plus -1ex \@minus -.2ex}
{2.3ex \@plus.2ex}
{\normalfont\Large\bfseries}}
\renewcommand\subsection{\@startsection{subsection}{2}{\z@}
{-3.25ex\@plus -1ex \@minus -.2ex}
{1.5ex \@plus.2ex}
{\normalfont\large\bfseries}}
\makeatother

\def\id{\mathbf{1}}

\def\fnsv{\vspace{3pt}}
\def\no{\nonumber}
\def\a{\alpha}

\def\d{\delta}
\def\e{\epsilon}

\def\h{\eta}

\def\z{\zeta}

\def\m{\mu}

\def\s{\sigma}

\def\c{\chi}

\def\hh{{\rm h}}
\def\coup{h}

\def\ket{\rangle}

\def\So{\EuScript{S}_{_0}}
\def\Po{\EuScript{P}_{_0}}
\def\tPo{\EuScript{\tilde P}_{_0}}

\def\pp{\textrm{p}}
\def\ee{\textrm{e}}
\def\mm{m}
\def\xpr{x'}
\def\ra{{\rm a}}
\def\qp{\theta_+}
\def\qm{\theta_-}
\def\qpm{\theta_\pm}
\def\qmp{\theta_\mp}

\def\khh{|\phi\phi\rangle}
\def\khs{|\phi\psi\rangle}
\def\ksh{|\psi\phi\rangle}
\def\kss{|\psi\psi\rangle}
\def\xx{{\rm x}}
\def\vv{{\rm v}}
\def\fl{f}
\def\combo{\frac{\sqrt{\xx^2}}{\xx}}
\def\combop{\frac{\xx'}{\sqrt{\xx'^2}}}
\newcommand{\alg}[1]{\mathfrak{#1}}
\newcommand{\comm}[2]{[#1,#2]}
\newcommand{\acomm}[2]{\{#1,#2\}}
\newcommand{\gen}[1]{\mathfrak{#1}}
\newcommand{\corr}[1]{{\color{black}#1}}
\newcommand{\corp}[1]{{\color{black}#1}}
\newcommand{\cortwo}[1]{{\color{black}#1}}

\begin{document}

\thispagestyle{empty}
\begin{flushright}\footnotesize\ttfamily
HU-EP-14/28
\\DMUS--MP--14/05
\end{flushright}
\vspace{2em}

\begin{center}

{\Large\bf \vspace{0.2cm}
{\color{black} Integrable S-matri\corr{ces}, massive and massless modes \\ and the $\mathbf{AdS_2 \times S^2}$ superstring}} 
\vspace{1.5cm}

\textrm{\large Ben Hoare,$^{a,}$\footnote{\texttt{ben.hoare@physik.hu-berlin.de}}
Antonio Pittelli,$^{b,}$\footnote{\texttt{a.pittelli@surrey.ac.uk}}
Alessandro Torrielli$^{b,}$\footnote{\texttt{a.torrielli@surrey.ac.uk}}}

\vspace{2em}

\vspace{1em}
\begingroup\itshape
${}^a$ Institut f\"ur Physik, Humboldt-Universit\"at zu Berlin,
\\ Newtonstra\ss e 15, 12489 Berlin, Germany.
\\\vspace{1em} ${}^b$ Department of Mathematics, University of Surrey,
\\ Guildford, GU2 7XH, UK.
\par\endgroup

\end{center}

\vspace{2em}

\begin{abstract}\noindent 
We derive the exact S-matrix for the 
\corr{scattering of {\color{black} particular representations of the centrally-extended \corp{$\alg{psu}(1|1)^2$}
Lie superalgebra, conjectured to be related to the massive modes of} the light-cone gauge string theory
on $AdS_2 \times S^2 \times T^6$.} The S-matrix 
consists of two copies of a centrally-extended \corp{$\alg{psu}(1|1)$} invariant
S-matrix and is in agreement with the tree-level result following from
perturbation theory. Although the overall factor is left unfixed, the
constraints following from crossing symmetry and unitarity are given. The
scattering involves long representations of the symmetry algebra, and the
relevant representation theory is studied in detail. We also discuss Yangian
symmetry and find it has a standard form \corr{for a particular limit of the
aforementioned representations. This has a natural interpretation as the
massless limit, and we investigate the corresponding limits of the massive
S-matrix.  Under the assumption that the massless modes of the light-cone gauge
string theory transform in these {\color{black} limiting} representations, the resulting
S-matrices would provide the building blocks for the full S-matrix.} Finally,
some brief comments are given on the Bethe ansatz.\end{abstract}

\newpage

\overfullrule=0pt
\parskip=2pt
\parindent=12pt
\headheight=0.0in \headsep=0.0in \topmargin=0.0in \oddsidemargin=0in

\vspace{-3cm}
\thispagestyle{empty}
\vspace{-1cm}

\tableofcontents

\setcounter{footnote}{0}

\section{Introduction}\label{secint}

The remarkable successes of integrability techniques in the study of the $AdS_5
\times S^5$ superstring \cite{rev,rev1} motivates the application of these methods
to other integrable string backgrounds with less supersymmetry \cite{sym}. In
this work we investigate the $AdS_2 \times S^2 \times T^6$ background supported
by Ramond-Ramond fluxes in Type II superstring theory, which preserves a
quarter of the supersymmetries. These can be found as the near-horizon limit of
various intersecting brane solutions of Type IIB supergravity, which are
related by T-duality \cite{ads2}. The dual \cite{adscft} should be a
one-dimensional CFT, and is understood to either be a superconformal
quantum-mechanical system or a chiral two-dimensional CFT \cite{dual}.

The $AdS_2 \times S^2$ part of the background can be written as a
Metsaev-Tseytlin \cite{Metsaev:1998it} type supercoset model \cite{sc} for
$PSU(1,1|2)/SO(1,1) \times SO(2)$. The algebra $\mathfrak{psu}(1,1|2)$ has a
$\mathbb{Z}_4$ automorphism and hence the supercoset model is classically
integrable via the same construction as for the $AdS_5 \times S^5$ case
\cite{Bena:2003wd}. While there exists a classical truncation of the
Green-Schwarz action \cite{Green:1984sg} for the $AdS_2 \times S^2 \times T^6$
geometry to the supercoset degrees of freedom, there is no $\kappa$-symmetry
gauge choice which decouples them from the remaining fermions
\cite{Sorokin:2011rr}. The integrability of the Green-Schwarz action for the
complete background has been demonstrated to quadratic order in fermions
\cite{Sorokin:2011rr,Cagnazzo:2011at}.

\

The aim of this paper is to use symmetries and integrability to construct exact
S-matrices for the scattering of the worldsheet excitations of the
decompactified light-cone gauge \cite{lcg} $AdS_2 \times S^2 \times T^6$
superstring. These S-matrices describe the scattering above the BMN vacuum
\cite{Berenstein:2002jq}, a point-like string moving at the speed of light on a
great circle of the two-sphere. The light-cone gauge-fixed Lagrangian
\cite{Murugan:2012mf,amsw} is in general rather complicated with the
interaction terms breaking two-dimensional Lorentz invariance. The quadratic
action is however Lorentz invariant and describes $2+2$ (bosons+fermions)
massive modes, the bosons of which are associated to the transverse directions
in $AdS_2 \times S^2$, and $6+6$ massless modes, associated to the $T^6$.

In the $AdS_5 \times S^5$ light-cone gauge-fixed theory all of the excitations
have equal non-vanishing mass and furthermore the symmetries completely fix the
S-matrix up to an overall phase \cite{beis0,conv,an}. Here the situation is more
similar to $AdS_3 \times S^3 \times T^4$ for which there are $4+4$ massive and
$4+4$ massless excitations. In this case the symmetries of the supercoset
leaving the BMN string invariant can be used to conjecture an exact S-matrix
for the scattering of the massive modes \cite{Borsato:2013qpa,HT} (see also the
review \cite{Sfond}). Following a similar approach we observe
that the subalgebra of the $\mathfrak{psu}(1,1|2)$ symmetry of the $AdS_2
\times S^2$ supercoset preserved by the BMN string is given by
$\mathfrak{psu}(1|1)^2\ltimes \mathbb{R}$. Relaxing the level-matching
condition we extend this algebra by two additional central extensions and
conjecture the exact S-matrix for the scattering of massive modes up to an
overall factor.

The resulting {\color{black} massive} S-matrix satisfies crossing symmetry \cite{Janik:2006dc} and is
unitarity so long as the overall factor satisfies the relevant identities. Here the
setup is more similar to the $AdS_5 \times S^5$ case as opposed to the
$AdS_3 \times S^3 \times T^4$ case for which there were multiple phases related
by crossing transformations \cite{ads3phase}. 
It was observed in \cite{amsw}
that the one-loop logarithms in the massive S-matrix for $AdS_2 \times S^2
\times T^6$ \corr{are consistent with the one-loop phase being related to the Hernandez-Lopez phase 
\cite{hl,bes}.}
Finally, the
near-BMN expansion of the exact result is consistent with perturbative
computations \cite{Murugan:2012mf,amsw,pert}.

\

While many features of the construction are similar to the $AdS_5 \times S^5$
and $AdS_3 \times S^3 \times T^4$ cases, there are some important differences.
In particular, unlike for the $AdS_5 \times S^5$ and $AdS_3 \times S^3 \times
T^4$ superstrings, the representations we are scattering turn out to be long
and hence there is no shortening condition to be interpreted as the dispersion
relation. An additional consequence is that the symmetries do not completely
fix the S-matrix up to a single overall factor, rather there is an additional
undetermined function that can be found by demanding the Yang-Baxter equation
is satisfied.  These properties are reminiscent of similar features seen for
the scattering of long representations of $\mathfrak{psu}(2|2) \ltimes
\mathbb{R}^3$ \cite{Arutyunov:2009pw} and also in the Pohlmeyer reduction of
strings on $AdS_2 \times S^2$ \cite{pr}.

The S-matrix has an accidental $U(1)$ symmetry under which the fermions are
charged, while the bosons are not. From the perspective of the complete $AdS_2
\times S^2 \times T^6$ superstring this $U(1)$ originates from the $T^6$
compact space \cite{amsw}. Furthermore, its presence appears to be important to
have any hope of applying a Bethe ansatz construction as it allows one to
define a pseudovacuum. A conjecture for a set of asymptotic Bethe ansatz
equations was given in \cite{Sorokin:2011rr}, however, due to the somewhat
involved structure of the S-matrix it is not clear how to derive them.

\

It is not currently known how the massless modes transform under the symmetry
group of the light-cone gauge-fixed theory, and therefore it is not possible to
completely determine the corresponding S-matrices. Furthermore, they may depend
on the choice of Type II background \cite{ads2} -- in the decompactified
light-cone gauge-fixed theory the $T^6$ formally has an $SO(6)$ symmetry,
however, this will be broken by the presence of Ramond-Ramond fluxes.
\corr{Initial investigations in this direction for a particular Type IIA
background were carried out in \cite{Murugan:2012mf,amsw}, in which case the
$SO(6)$ is broken to $U(3)$.  However, different backgrounds related by T-duality
will naively lead to different subgroups \cite{ads2}.} 

Here we take an alternative \corr{(partial)} approach \corr{to the question of massless
modes} motivated by the recent explicit
computation of the light-cone gauge symmetry algebra for the $AdS_3 \times S^3
\times T^4$ superstring \cite{BogdanLatest}, the $AdS_5 \times S^5$ version of
which was constructed in \cite{Arutyunov:2006ak}.  Under the assumption that a
similar outcome occurs for the $AdS_2 \times S^2 \times T^6$ superstring one
may expect the massless modes to transform in representations of
$\alg{psu}(1|1)^2\ltimes \mathbb{R}^3$. {\color{black} We further rely on the fact that the two-dimensional modules we work with are rather general in their parametrization, and assume that the massless representations take the same form as the massive ones, provided one sends the mass parameter $m$ to zero. Upon adopting these assumptions,} the S-matrices describing
their scattering should be built from the massless limits (one massless and one
massive or two massless particles) of the massive S-matrix. \corr{How these
building blocks are precisely put together and the corresponding overall number
of undetermined phases (of which there may be many) will depend on the complete
symmetry of the light-cone gauge-fixed backgrounds (including the subgroup of
$SO(6)$ preserved by the fluxes), an analysis that we leave for future work.} 

\

The structure of the paper is as follows. In section \ref{sec1} we describe the
near-BMN symmetry algebra and investigate its representation theory. This
symmetry is then used in section \ref{sec2} to determine the exact S-matrix up
to an overall phase. We determine the constraints that the phase should satisfy
for crossing symmetry and unitarity and compare with perturbation theory. In
section \ref{Yangian} we discuss when this symmetry can be extended to a
Yangian, finding that it can be done in the standard form for the massless
case. Using this Yangian symmetry in section \ref{massless} we then construct
the massless version of the S-matrix and briefly explore the notions of
crossing symmetry and unitarity in this limit. In section \ref{secu1} we give
some initial considerations of the algebraic Bethe ansatz, noting in particular
the existence of a pseudovacuum, and we conclude in section \ref{comments} with
some comments.

\section{Symmetry for massive modes of \texorpdfstring{$\mathbf{AdS_2 \times S^2}$}{AdS2 x S2}}\label{sec1}

The BMN light-cone gauge $AdS_2 \times S^2 \times T^6$ superstring action
describes $2+2$ massive and $6+6$ massless modes. The algebra
underlying the scattering of the massive modes is expected to be
$\mathfrak{psu}(1|1)^2 \ltimes \mathbb{R}^3$, which is found by considering the
subalgebra of $\mathfrak{psu}(1,1|2)$ that is preserved by the BMN geodesic.
{\color{black} We expect two additional central extensions to appear}, by analogy with the $AdS_5 \times
S^5$ case, in the decompactification limit and relaxing the level-matching
condition. 

{\color{black} Although a full off-shell analysis, as in  \cite{Arutyunov:2006ak,BogdanLatest}, would be necessary (and is planned \corp{for 
future work)} 
to confirm the nature of the central extensions,
\corp{in this paper we construct the massive S-matrix on the basis of certain assumptions.}
\corp{The} first assumption is the analogy with higher dimensional AdS/CFT integrable \corp{systems}, 
\corp{and in particular the} way the central extensions manifest themselves. 
\corp{This assumption is also motivated by} 
perturbation theory, in particular 
\corp{the symmetry algebra being given
by} two copies of a centrally-extended algebra with the centres 
identified, seems to be suggested, for instance, by the work of \cite{amsw}. 
\corp{The second} 
 crucial assumption is integrability itself. On the one hand, integrability should 
work \corp{to complete} 
the perturbative results into the structure of classified representations of superalgebras. 
On the other hand, it should maintain the tree-level factorized form of the S-matrix at higher string loops. 

With these assumptions in mind, we will nevertheless \corp{pursue} a broad approach and 
explore the most general central extension based on the available kinematical algebra.
We denote} the massive boson associated to the transverse direction of
$S^2$ as $y$ and the corresponding boson for $AdS_2$ as $z$. The two massive
fermions will be represented as two real Grassmann fields $\zeta$ and $\chi$.
We can then formally define the following tensor product states
\begin{equation}\begin{split}\label{bhtens}
\left|y\right> = \left|\phi\right> \otimes \left|\phi\right> \ , \qquad &
\left|z\right> = \left|\psi\right> \otimes \left|\psi\right> \ , \\
\left|\zeta\right> = \left|\phi\right> \otimes\left|\psi\right> \ , \qquad &
\left|\chi\right> = \left|\psi\right> \otimes \left|\phi\right> \ ,
\end{split}\end{equation}
where $\phi$ is bosonic and $\psi$ is fermionic, such that we expect one of the
factors of $\mathfrak{psu}(1|1)$ to act on each of the two entries.
Furthermore, as a consequence of the form of the symmetry algebra and the
integrability of the theory \cite{Sorokin:2011rr,Murugan:2012mf} we expect that
the S-matrix for $y$, $z$, $\zeta$ and $\chi$ can be constructed as a graded
tensor product of an S-matrix for $\phi$ and $\psi$, with each factor S-matrix
invariant under the symmetry $\mathfrak{psu}(1|1) \ltimes \mathbb{R}^3$.

In this section we will construct the relevant massive representation of
$\mathfrak{psu}(1|1) \ltimes \mathbb{R}^3$. This representation has an obvious
massless limit, and, by analogy with the construction for $AdS_3 \times S^3
\times T^4$ \cite{BogdanLatest}, one may expect the massless modes to also
transform in representations of $\mathfrak{psu}(1|1) \ltimes \mathbb{R}^3$ in
the light-cone gauge-fixed theory. The massless limit is discussed in detail
in section \ref{massless}.

Let us also briefly mention that there is an additional $U(1)$ outer
automorphism symmetry \cite{amsw} of the S-matrix \eqref{redu}, under which the
$\mathfrak{psu}(1|1)$ factors transform in the vector representation. The
origin of this $U(1)$ symmetry is the $T^6$ compact space that is required for
a consistent 10-d superstring theory. Under this symmetry $(\z,\chi)^T$ also
transforms as a vector, while the bosons are uncharged. It is worth noting that
taking the tensor product of two copies of any S-matrix for $\phi$ and $\psi$
\corr{preserving the value of $(-1)^F$, where $F$ is the fermion number operator,} 
we find that the $U(1)$ symmetry is present so long as a certain quadratic
relation between the parametrizing functions is satisfied (see appendix
\ref{appu1}). In the case of interest, this quadratic identity turns out to be
true just from demanding invariance under the $\mathfrak{psu}(1|1) \ltimes
\mathbb{R}^3$ symmetry and satisfaction of the Yang-Baxter equation. The $U(1)$
does not act in a well-defined way on the individual factor S-matrices and
hence for now we will ignore it. We will reconsider it in section \ref{secu1},
where it will play a role in defining a pseudovacuum, an important first step
in the algebraic Bethe ansatz.

\subsection{The \texorpdfstring{$\mathfrak{gl}(1|1)$}{gl(1|1)} Lie superalgebra and its representations}\label{gl1}

\begin{figure}[ht]
    \centerline{\includegraphics[scale=0.3,viewport = 0 0 16cm 18.1cm]{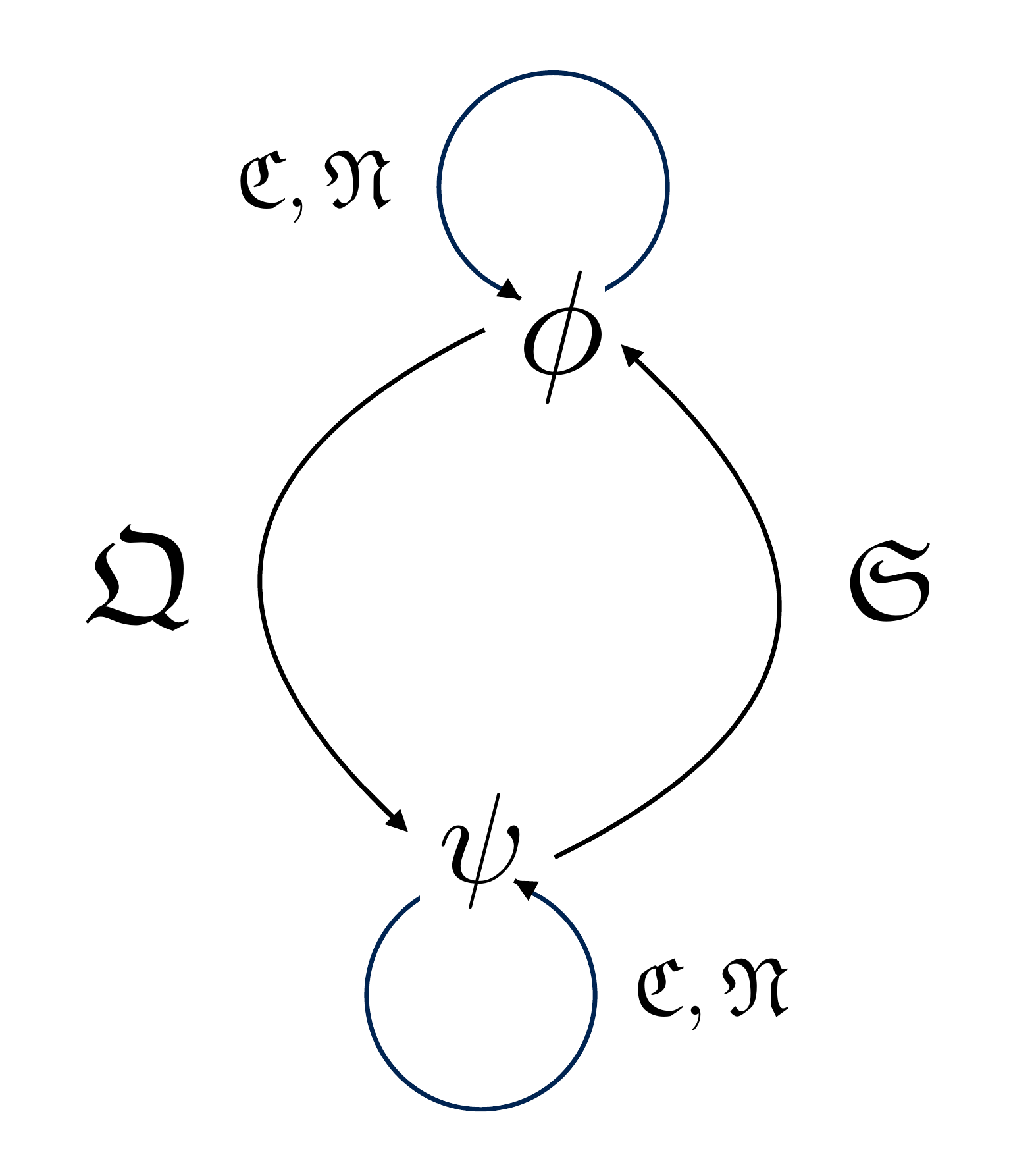}}
   \caption{The Kac module $\langle C,\nu\rangle$.}
\label{fig:fig1}
\end{figure}

Let us start by summarizing the relevant information from
\cite{GotzQuellaSchomerus} regarding the Lie superalgebra $\mathfrak{gl}(1|1)$
and its representations. There are two bosonic generators $\mathfrak N$ and
$\mathfrak C$, with $\mathfrak C$ central, and two fermionic generators
$\mathfrak Q$ and $\mathfrak S$. The commutation relations read
\begin{equation}\label{com0}
[\mathfrak N , \,\mathfrak Q] = - \mathfrak Q \ , \qquad
[\mathfrak N , \,\mathfrak S] = \mathfrak S \ , \qquad
\{\mathfrak Q, \,\mathfrak S \} = 2 \mathfrak C \ .
\end{equation}
The typical (long) irreps are the 2-dimensional {\it Kac modules} $\langle C,
\nu\rangle$, defined by the following non-zero entries on a boson-fermion
$(|\phi\rangle, \, |\psi\rangle)$ pair of states:
\begin{eqnarray}
&& \mathfrak Q\left|\phi\right>=\left|\psi\right> \ , \qquad
\mathfrak S\left|\psi\right>=2 \, C \left|\phi\right> \ , \qquad
\mathfrak N\left|\phi\right>=\nu\left|\phi\right> \ , \qquad
\mathfrak N\left|\psi\right>=(\nu-1)\left|\psi\right> \ , \nonumber
\\
&& \mathfrak C\left|\Phi\right>= C\left|\Phi \right> \ \qquad
\forall \, \, \, |\Phi\rangle \, \in \,\{ |\phi\rangle, \, |\psi\rangle\}, \qquad
C, \nu \, \in \, \mathbb{C}, \qquad C \neq 0. \label{undefa}
\end{eqnarray}
{\color{black} We have summarized the generator action in figure \ref{fig:fig1}.} 
As long as $C \neq 0$, this module is isomorphic to the {\it anti-Kac module}
$\overline{\langle C, \nu\rangle}$
\begin{eqnarray}
&& \mathfrak Q\left|\psi\right>=2 \, C \left|\phi\right> \ , \qquad
\mathfrak S\left|\phi\right>=\left|\psi\right> \ , \qquad
\mathfrak N\left|\phi\right>=(\nu-1)\left|\phi\right> \ , \qquad
\mathfrak N\left|\psi\right>=\nu\left|\psi\right> \ , \nonumber
\\
&& \mathfrak C\left|\Phi\right>= C\left|\Phi \right> \ \qquad
\forall \, \, \, |\Phi\rangle \, \in \, \{|\phi\rangle, \, |\psi\rangle\}, \qquad
C, \nu \, \in \, \mathbb{C}, \qquad C \neq 0. \label{undefb}
\end{eqnarray}
However, if $C=0$, the two modules are not isomorphic and they are no longer
irreducible. Rather they become reducible but indecomposable.

To elucidate further we introduce the 1-dimensional modules $\langle \mu
\rangle$, which form the atypical (short) irreps of $\mathfrak{gl}(1|1)$.
These irreps are characterized by the vanishing of all generators except
$\mathfrak N$, which acts with eigenvalue $\mu$. We then see that for the Kac
module, $\langle 0, \nu\rangle$, the fermion $|\psi\rangle$ spans a
sub-representation $\langle \nu-1\rangle$, and the indecomposable is denoted as
\begin{eqnarray}
\langle \nu-1 \rangle \, \longleftarrow \langle \nu \rangle \ .
\end{eqnarray}
The anti-Kac module $\overline{\langle 0, \nu\rangle}$ is also reducible but
indecomposable and is denoted as
\begin{eqnarray}
\langle \nu-1 \rangle \, \longrightarrow \langle \nu \rangle\ ,
\end{eqnarray}
with the fermion $|\psi\rangle$ once again spanning the sub-representation
$\langle \nu \rangle$. This indecomposable is {\it not} isomorphic to $\langle
0, \nu\rangle$. Let us mention that modding out the indecomposable
representations by their sub-representations one obtains the {\it factor}
representations, which in this case are isomorphic to the short 1-dimensional
$\langle \mu \rangle$ modules and are spanned by the boson $|\phi\rangle$.

If we take the tensor product of two typical modules, we get
\begin{eqnarray}
&& \langle C_1, \nu_1 \rangle \otimes \langle C_2, \nu_2 \rangle \,
= \, \langle C_1 + C_2, \nu_1 + \nu_2 - 1 \rangle \oplus \langle C_1 + C_2, \nu_1 + \nu_2 \rangle \qquad
\mbox{if} \, \, \, C_1 + C_2 \neq 0\ , \no
\\
&&\langle C_1, \nu_1 \rangle \otimes \langle - C_1, \nu_2 \rangle \,
= \, P_{\nu_1 + \nu_2}\ ,
\end{eqnarray}
where $P_\nu$ is the so-called {\it projective module}
\begin{eqnarray}
\langle \nu \rangle \, \longrightarrow \, \langle \nu + 1 \rangle \oplus \langle \nu - 1 \rangle \, \longrightarrow \, \langle \nu \rangle\ ,
\end{eqnarray}
on which $\mathfrak C$ acts identically as zero. The rightmost 1-dimensional
short sub-module $\langle \nu \rangle$ is known as the {\it socle} of $P_\nu$.

Since $\mathfrak N$ does not appear on the r.h.s. of the commutation relations,
the algebra $\mathfrak{gl}(1|1)$ has a non-trivial ideal generated by
$\mathfrak Q$, $\mathfrak S$ and $\mathfrak C$. This ideal is the superalgebra
$\mathfrak{sl}(1|1)$. Furthermore, this algebra is also not simple, as
$\mathfrak C$, being central, is a non-trivial ideal. Additionally modding out
$\mathfrak C$ gives the algebra $\mathfrak{psl}(1|1)$, which is still not
simple, as the two remaining anti-commuting fermionic generators each form a
separate ideal. The fact that $\mathfrak{psl}(1|1)$ is not simple sets this
algebra outside the classification of the possible central extensions of basic
classical Lie superalgebras presented in \cite{IoharaKoga}.

\subsection{The centrally-extended \texorpdfstring{$\mathfrak{psu}(1|1)$}{psu(1|1)} Lie superalgebra}\label{massi}

\begin{figure}[ht]
    \centerline{\includegraphics[scale=0.2,viewport = 0 0 36.1cm 27.1cm]{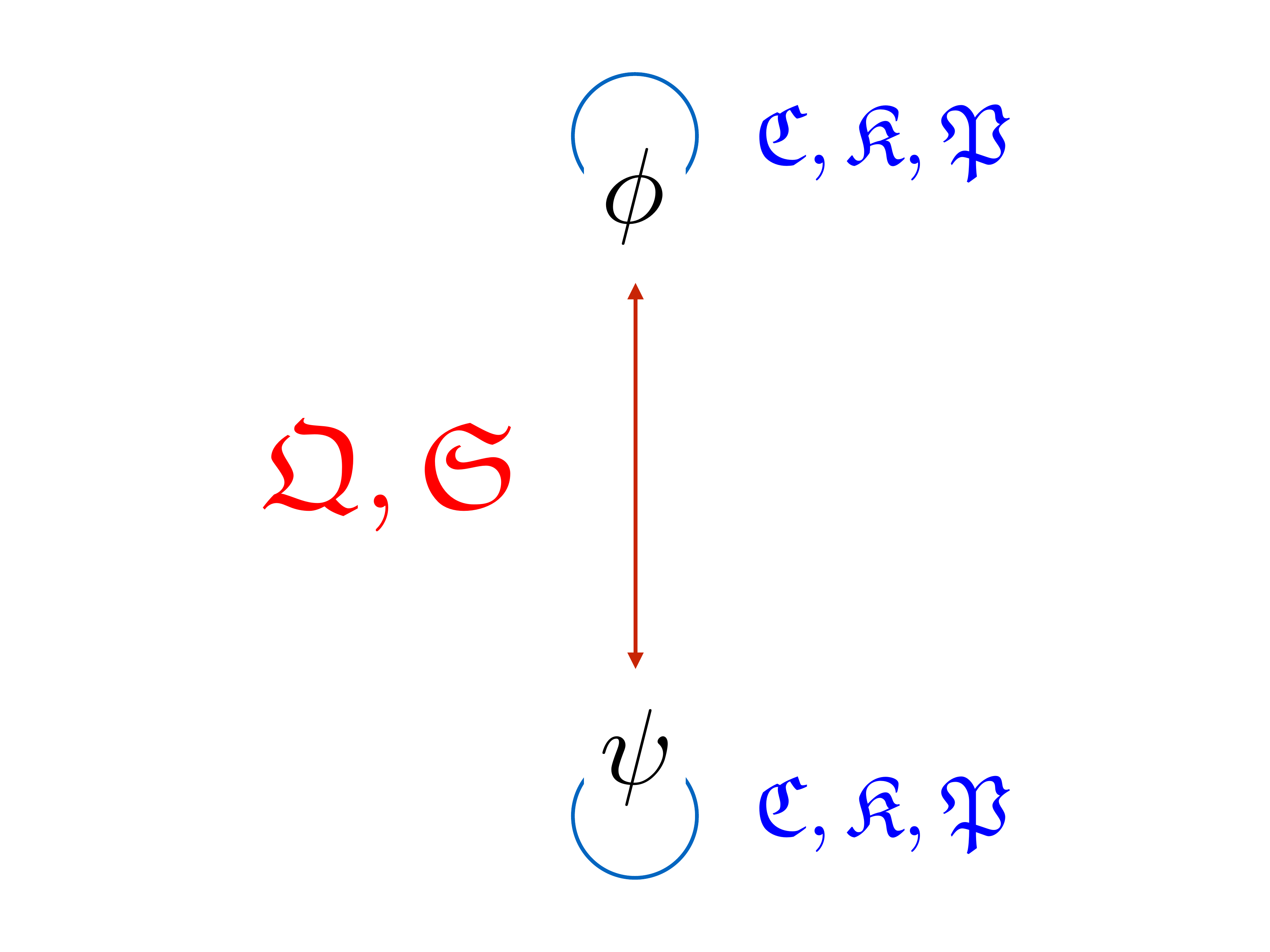}}
   \caption{The 2-dimensional module of the centrally-extended algebra (several lines are superimposed).}
\label{fig:fig2}
\end{figure}

We are now ready to introduce the centrally-extended version of the algebra we
discussed {\color{black} above, which, as anticipated by the discussion at the beginning of section \ref{sec1}, we conjecture to be} relevant for the scattering of the
massive modes of the $AdS_2 \times S^2 \times T^6$ superstring. The algebra
$\mathfrak{psu}(1|1) \ltimes \mathbb R^3$ is defined by the commutation
relations
\begin{equation}\label{com}
\{\mathfrak Q , \,\mathfrak Q\} =2 \mathfrak P \ , \qquad
\{\mathfrak S, \,\mathfrak S\} = 2 \mathfrak K \ , \qquad
\{\mathfrak Q, \,\mathfrak S \} = 2 \mathfrak C \ .
\end{equation}
The states $|\phi\rangle$ and $|\psi\rangle$, introduced in \eqref{bhtens},
then transform in the following representation:
\begin{align}
& \mathfrak Q\left|\phi\right>=a\left|\psi\right> \ , & & & &
\mathfrak Q\left|\psi\right>=b \left|\phi\right> \ , & & & &
\mathfrak S\left|\phi\right>=c\left|\psi\right> \ , & & & &
\mathfrak S\left|\psi\right>=d\left|\phi\right> \ , \nonumber
\\
& \mathfrak C\left|\Phi\right>= C\left|\Phi \right> \ , & & & &
\mathfrak P\left|\Phi\right>=P \left|\Phi \right> \ , & & & &
\mathfrak K\left|\Phi\right>= K \left|\Phi \right> \ . \label{undef}
\end{align}
Here $a$, $b$, $c$, $d$, $C$, $P$ and $K$ are the representation parameters
that will eventually be functions of the energy and momentum of the states. For
the supersymmetry algebra to close the following conditions should be satisfied
\begin{equation}\begin{split}\label{abcdCPK}
a b = P \ , \qquad c d = K \ , \qquad & a d + b c = 2 C \ .
\end{split}\end{equation}
This representation corresponds to the typical (long) Kac module $\langle C,
\nu\rangle$ discussed in the previous section. {\color{black} We have summarized the generator action in figure \ref{fig:fig2}.} 
We will be interested in a
particular real form of the algebra \eqref{com}, which is given by
\begin{equation}\label{hermi}
\mathfrak{Q}^\dagger =\mathfrak{S}\ , \qquad
\mathfrak{P}^\dagger = \mathfrak{K} \ , \qquad
\mathfrak{C}^\dagger =\mathfrak{C}\ .
\end{equation}
These relations further constrain the representation parameters as follows
\begin{equation}\label{rc}
a^* = d \ , \qquad b^* = c \ , \qquad C^* = C \ , \qquad P^* = K \ .
\end{equation}

The closure conditions \eqref{abcdCPK} imply that
\begin{equation}\label{short}
C^2 = \frac{(ad-bc)^2}4 + P K \ .
\end{equation}
Unlike the $AdS_5 \times S^5$ case, with the larger symmetry algebra
$\mathfrak{psu}(2|2)^2 \ltimes \mathbb R^3$, here we are scattering long
representations and hence there is no shortening condition -- that is, $ad -
bc$ is free to take any value, \cortwo{which we denote
\begin{equation}
\mm\equiv ad-bc \ .
\end{equation}
The reality conditions \eqref{rc} imply that $\mm$ is real. 
From \eqref{short} we then have
\begin{equation}
(C+\frac{m}{2})(C-\frac{m}{2}) = PK > 0 \ ,
\end{equation}
also as a consequence of the reality conditions \eqref{rc}.
Motivated by the fact that $C$ will later be associated to an
energy, we will take it to be positive. {\color{black} However, let us point out that the algebraic analysis we perform in this paper is largely insensitive to this choice, \cortwo{and hence it} does not represent a loss of generality. 
If we make this positivity assumption, it} immediately follows that both $(C+\frac{m}2)$ and $(C -\frac{m}{2})$
are also positive.} 
{\color{black} The analogy with the higher dimensional AdS/CFT cases 
\corp{suggests that we should associate \cortwo{(the absolute value of)} $m$ with the}
mass of the scattering particle.}
Later it will be useful to solve the set of equations \eqref{abcdCPK} for
$a$, $b$, $c$ and $d$ in terms of $\mm$, $C$, $P$ and $K$
\begin{align}\nonumber
a =\, & \a \,e^{-\frac{i\pi}4} \big(C + \frac{\mm}2\big)^{\frac12} \ , &\
b =\, & \a^{-1} e^{\frac{i\pi}4} \big(C + \frac{\mm}2\big)^{-\frac12}P \ , \\
c =\, & \a\, e^{-\frac{i\pi}4} \big(C + \frac{\mm}2\big)^{-\frac12}K \ , &
d =\, & \a^{-1}e^{\frac{i\pi}4} \big(C + \frac{\mm}2\big)^{\frac12} \ . \label{ad}
\end{align}
Here $\a$ is a phase parametrizing the normalization of the fermionic states
with respect to the bosonic states and can be a function of the central
extensions.

\

To define the action of this symmetry on the two-particle states we need to
introduce the coproduct
\begin{align}
\Delta(\mathfrak Q) & = \mathfrak Q \otimes \id + \mathfrak U \otimes \mathfrak Q \ , & \nonumber & &
\Delta(\mathfrak S) & = \mathfrak S \otimes \id + \mathfrak U^{-1} \otimes \mathfrak S \ ,
\\
\Delta(\mathfrak P) & = \mathfrak P \otimes \id + \mathfrak{U}^2 \otimes \mathfrak P \ , &
\Delta(\mathfrak C) & = \mathfrak C \otimes \id + \id \otimes \mathfrak C \ , &
\Delta(\mathfrak K) & = \mathfrak K \otimes \id + \mathfrak{U}^{-2} \otimes \mathfrak K \ ,\label{coproduct}
\end{align}
and the opposite coproduct, defined as
\begin{equation}
\Delta^{\text{op}}(\mathfrak J) = \mathcal{P}\,\Delta(\mathfrak J)\ ,
\end{equation}
where $\mathfrak J$ is an arbitrary abstract generator (prior to considering
a representation), and $\mathcal{P}$ defines the graded permutation of the tensor product.

The coproduct differs from the trivial one by the introduction of a new abelian
generator $\mathfrak U$, with $\Delta(\mathfrak U) = \mathfrak U \otimes
\mathfrak U$ \cite{tor}. This is done according to a $\mathbb{Z}$-grading of
the algebra, whereby the charges $-2,-1,1$ and $2$ are associated to the
generators $\mathfrak K$, $\mathfrak S$, $\mathfrak Q$ and $\mathfrak P$
respectively, while $\mathfrak C$ remains uncharged. The action of $\mathfrak
U$ on the single-particle states is given by
\begin{equation}
\mathfrak U\left|\phi\right> = U\left|\phi\right> \ , \qquad\qquad
\mathfrak U\left|\psi\right> = U\left|\psi\right> \ .\label{uu}
\end{equation}
This braiding allows for the existence of a non-trivial S-matrix.

One important consequence of the non-trivial braiding \eqref{coproduct} is that
it leads to a constraint between $U$ and the eigenvalues of the central
charges. This follows from the requirement that, to admit an S-matrix, the
coproduct of any central element should be equal to its
opposite.\footnote{\label{foo10}If $\Delta(\gen{c})$ is central, then
\begin{equation*}
\Delta^{op}(\gen{c}) \, R \, = \, R \, \Delta(\gen{c}) = \Delta(\gen{c}) \, R\ ,
\end{equation*}
which, for an invertible R-matrix, necessarily implies
$\Delta^{op}(\gen{c})=\Delta(\gen{c})$. This is expressed by saying that the
coproduct of $\gen{c}$ is {\it co-commutative}.\fnsv} This implies
\begin{equation}
\mathfrak P \propto (1 - \mathfrak U^2) \ ,\qquad \qquad \mathfrak K \propto (1 - \mathfrak U^{-2}) \ . \label{pp}
\end{equation}
We fix the normalization of $\mathfrak P$ relative to $\mathfrak K$ by taking
both constants of proportionality to be equal to $\tfrac12 \hh$ where the
reality conditions \eqref{rc} require that $\hh$ is real.\footnote{The reality
conditions \eqref{rc} do allow for the introduction of an additional phase into
the constants of proportionality, {\em i.e.} $\tfrac12 \hh e^{i\varphi}$ and
$\tfrac12 \hh e^{-i\varphi}$. However, this phase does not appear in the
S-matrix and thus we set $\varphi=0$.\fnsv} The parameter $\hh$ is a coupling
constant and eventually should be fixed in terms of the string tension, which
we will return to in section \ref{secpert}. Acting on the single-particle
states then gives us the relations
\begin{equation}\label{pku}
P = \frac \hh2 \, (1 - U^2) \ ,\qquad \qquad K = \frac \hh2 \, (1 - U^{-2}) \ ,
\end{equation}
where $U$ should satisfy, as a consequence of \eqref{rc}, the following reality
condition
\begin{equation}\label{urc}
U^* = U^{-1} \ .
\end{equation}
The relation \eqref{short} in terms of $C$, $U$ and $\mm$ is then given by
\begin{equation}\label{apii}
C^2 = \frac{\mm^2-\hh^2(U-U^{-1})^2}4\ .
\end{equation}
While this is a single equation for three undetermined parameters, we will
later still attempt to interpret it as a dispersion relation with $C$, $U$
and $\mm$ defined in terms of just two kinematic variables, the energy and
momentum. These precise definitions are not fixed by symmetry considerations,
and hence should be found from direct string computations.

\

It is now useful to introduce the Zhukovsky variables $x^\pm$, in terms of
which we will write the S-matrix, in place of the central extensions, $C$ and
$U$. These are defined as \cite{beis0,beis1}
\begin{equation} \label{xx}
U^2 = \frac{x^+}{x^-} \ , \qquad\qquad 2C + \mm = i \hh (x^- - x^+) \ ,
\end{equation}
In these variables the dispersion relation \eqref{apii} takes the following
familiar form
\begin{equation}
x^+ + \frac{1}{x^+} -x^- - \frac{1}{x^-} = \frac{2i\mm}{\hh} \ . \label{yy}
\end{equation}
The representation parameters $a$, $b$, $c$ and $d$ in \eqref{ad} and
\eqref{bep} are then given by
\begin{align}
& a =\, \a \,e^{-\frac{i\pi}4} \sqrt[4]{\frac{x^+}{x^-}}\sqrt{\frac{\hh}2}\ \h \ , \qquad
&& b =\, \a^{-1} e^{-\frac{i\pi}4} \sqrt[4]{\frac{x^-}{x^+}}\sqrt{\frac{\hh}2} \ \frac{\h}{x^-} \ , \no \\
& c =\, \a\, e^{\frac{i\pi}4} \sqrt[4]{\frac{x^+}{x^-}} \sqrt{\frac{\hh}2} \ \frac{\h}{x^+} \ , \qquad
&& d =\, \a^{-1}e^{\frac{i\pi}4} \sqrt[4]{\frac{x^-}{x^+}} \sqrt{\frac{\hh}2}\ \h \ , \label{da}
\end{align}
where
\begin{equation}\label{ett}
\h \equiv \sqrt{i(x^- - x^+)} \ .
\end{equation}
Here we clearly see that the advantage of these variables is that the
parameters $a$, $b$, $c$ and $d$ do not depend on $\mm$ and hence, written as a
function of $x^\pm$ and $m$, neither will the S-matrix. Finally, let us note
that for the reality conditions \eqref{rc} we have the usual $(x^\pm)^* =
x^\mp$.

\

We could also eliminate the central extensions, $C$ and $U$, in terms of two
variables that will later be identified with the energy and momentum. Motivated
by the $AdS_5 \times S^5$ case we write
\begin{equation}\label{cuep}
C = \frac{\ee}{2} \ ,\qquad \qquad U = e^{\frac i2 \pp} \ ,
\end{equation}
where $\ee$ is the energy and $\pp$ is the spatial momentum. \corr{While the 
identification of $\ee$ with the energy and $\pp$ with the spatial momentum is
{\color{black} at present} only motivated by analogy with the $AdS_5 \times S^5$ case, a
posteriori it will be further justified by matching with perturbative results
in section \ref{secpert}.} Solving for $x^\pm$ in terms of $\ee$ and $\pp$ we find
\begin{equation}\begin{split}
x^\pm = r\, U^{\pm 1} \ , \qquad
r= \frac{ \ee + \mm }{2 \hh\,\sin\frac {\pp}2} =\frac{2\hh\,\sin\frac{\pp}2}{\ee - \mm }\ , \qquad
U = e^{\frac{i\pp}{2}} \ .
\label{4444}
\end{split}\end{equation}
Using \eqref{pku} and \eqref{cuep} we can substitute in for $C$, $P$ and $K$
in terms of the energy and the momentum in \eqref{apii} to find the following
familiar dispersion relation
\begin{equation}\label{apiii}
\ee^2 = \mm^2 + 4 \, \hh^2 \, \sin^2\frac{\pp}2 \ .
\end{equation}
It is important to emphasize that here $\mm$ is algebraically a free parameter.
However, for \eqref{apiii} to really be interpreted as a dispersion relation
$m$ should be fixed by the spectral analysis of the theory. In terms of the
energy and the momentum the representation parameters $a$, $b$, $c$ and $d$
\eqref{ad} are given by
\begin{align}\nonumber
a =\, & \frac{\a \,e^{\frac{i\pp}{4}-\frac{i\pi}4}}{\sqrt2} \sqrt{\ee+\mm} \ , &\
b =\, & \frac{\a^{-1} e^{-\frac{i\pp}{4}+\frac{i\pi}4}}{\sqrt2} \frac{\hh(1-e^{i\pp})}{\sqrt{\ee+\mm}} \ , \\
c =\, & \frac{\a\, e^{\frac{i\pp}{4}-\frac{i\pi}4}}{\sqrt2} \frac{\hh(1-e^{-i\pp})}{\sqrt{\ee+\mm}} \ , &
d =\, & \frac{\a^{-1}e^{-\frac{i\pp}{4}+\frac{i\pi}4}}{\sqrt 2} \sqrt{\ee+\mm} \ . \label{bep}
\end{align}

In the $AdS_5 \times S^5$ and $AdS_3 \times S^3\times M^4$ models, the choice
of the phase factor $\a$ that is appropriate for the light-cone gauge-fixed
string theory is
\begin{equation}\label{nca}
\a = 1\ .
\end{equation}
As we will see, this is also a natural choice for $\a$ in the $AdS_2 \times
S^2$ theory.

\subsection{Tensor product of irreps and scattering theory}\label{32}

In this section we consider the tensor product of two of the irreps we
discussed in the previous section, with the aim of constructing the relevant
scattering theory. In particular, we want to investigate the persistence of the phenomenon
observed for $\alg{gl}(1|1)$ modules in section \ref{gl1}, namely complete
reducibility of the tensor product of two 2-dimensional irreps, for {\it
generic} values of the momenta, into two 2-dimensional irreps of the same type.

Let us proceed by constructing a 4-dimensional representation of the algebra
\eqref{com}. To do this we start with the bosonic state
\begin{equation}
|w_0\rangle \ .
\end{equation}
Let us assume that the action of the central elements on this state is given by
\begin{equation}
(\alg{P},\alg{K},\alg{C})|w_0\rangle = (P,K,C)|w_0\rangle \ .
\end{equation}
This assumption will be justified by the concrete example we will consider
later in our treatment of the scattering theory. We can then construct two
more states by considering the action of $\alg{Q}$ and $\alg{S}$
\begin{equation}
|w_1\rangle \equiv \alg{Q}|w_0\rangle \ , \qquad
|\tilde w_1\rangle \equiv \alg{S}|w_0\rangle \ .
\end{equation}
The action of the central elements on these new states is then easily seen to
be given by
\begin{equation}
(\alg{P},\alg{K},\alg{C})|w_1\rangle = (P,K,C)|w_1\rangle \ , \qquad
(\alg{P},\alg{K},\alg{C})|\tilde w_1\rangle = (P,K,C)|\tilde w_1\rangle \ .
\end{equation}
We can then look at the action of $\alg{Q}$ and $\alg{S}$ on
$|w_1\rangle$ and $|\tilde w_1\rangle$
\begin{equation}\begin{split}
\alg{Q}|w_1\rangle = & P|w_0\rangle \ , \qquad
\alg{Q}|\tilde w_1\rangle = C|w_0\rangle + \frac12[\alg{Q},\alg{S}]|w_0\rangle \ ,
\\
\alg{S}|\tilde w_1\rangle = & K|w_0\rangle \ , \qquad
\alg{S}|w_1\rangle = C|w_0\rangle - \frac12[\alg{Q},\alg{S}]|w_0\rangle \ .
\end{split}\end{equation}
Here we see that we have generated one additional new state
\begin{equation}
|\tilde w_0\rangle \equiv \frac 1{M} [\alg{Q},\alg{S}]|w_0\rangle \ ,
\end{equation}
where we have chosen a normalization depending on
\begin{equation}\label{mdef}
M \equiv 2 \sqrt{C^2 - PK}\ .
\end{equation}
Given the real form we are interested in, see eq.~\eqref{hermi}, and the
assumption that $C^2>PK$, or equivalently that $M$ is real and non-zero (we
will briefly discuss the case when $M$ vanishes at the end of this section),
the above normalization implies that $|\tilde w_0\rangle$ has the same norm as
$|w_0\rangle$. Therefore, the action of $\alg{Q}$ and $\alg{S}$ on
$|w_1\rangle$ and $|\tilde w_1\rangle$ is given by
\begin{equation}\begin{split}
\alg{Q}|w_1\rangle = & P|w_0\rangle \ , \qquad
\alg{Q}|\tilde w_1\rangle = C|w_0\rangle + \frac{M}2|\tilde w_0\rangle \ ,
\\
\alg{S}|\tilde w_1\rangle = & K|w_0\rangle \ , \qquad
\alg{S}|w_1\rangle = C|w_0\rangle - \frac{M}2|\tilde w_0\rangle \ .
\end{split}\end{equation}
Again it is clear that the action of the central elements on $|\tilde
w_0\rangle$ is given by
\begin{equation}
(\alg{P},\alg{K},\alg{C})|\tilde w_0\rangle = (P,K,C)|\tilde w_0\rangle \ .
\end{equation}
Finally, the action of $\alg{Q}$ and $\alg{S}$ on $|\tilde w_0\rangle$ is given by
\begin{equation}
\alg{Q}|\tilde w_0\rangle = \frac{2P}{M}|\tilde w_1\rangle - \frac{2C}{M} |w_1\rangle \ , \qquad
\alg{S}|\tilde w_0\rangle = -\frac{2K}{M} |w_1\rangle + \frac{2C}{M} |\tilde w_1\rangle \ .
\end{equation}
Therefore, in summary, we have constructed the following 4-dimensional
representation:
\begin{equation*}
(\alg{P},\alg{K},\alg{C})|\Phi\rangle = (P,K,C)|\Phi\rangle \ , \qquad
\forall \ |\Phi\rangle \, \in \, \{|w_0\rangle,|w_1\rangle,|\tilde w_1\rangle,|\tilde w_0\rangle)\} \ ,
\end{equation*}
\begin{align}
\alg{Q}|w_0\rangle = &\ |w_1\rangle \ , & \alg{S}|w_0\rangle = &\ |\tilde w_1\rangle \ ,\no
\\
\alg{Q}|w_1\rangle = &\ P|w_0\rangle \ , & \alg{S}|\tilde w_1\rangle = &\ K |w_0\rangle \ ,\no
\\
\alg{Q}|\tilde w_1\rangle = &\ C |w_0\rangle + \frac{M}2 |\tilde w_0\rangle \ , & \alg{S}|w_1\rangle = &\ C |w_0\rangle - \frac{M}2 |\tilde w_0\rangle \ ,\no
\\
\alg{Q}|\tilde w_0\rangle = &\ \frac {2P}{M} |\tilde w_1\rangle - \frac {2C}{M} |w_1\rangle \ , & \alg{S}|\tilde w_0\rangle = &\ - \frac {2K}{M} |w_1\rangle + \frac {2C}{M} |\tilde w_1\rangle \ .
\end{align}

\begin{figure}[ht]
    \centerline{\includegraphics[scale=0.3,viewport = 0 0 36.1cm 27.1cm]{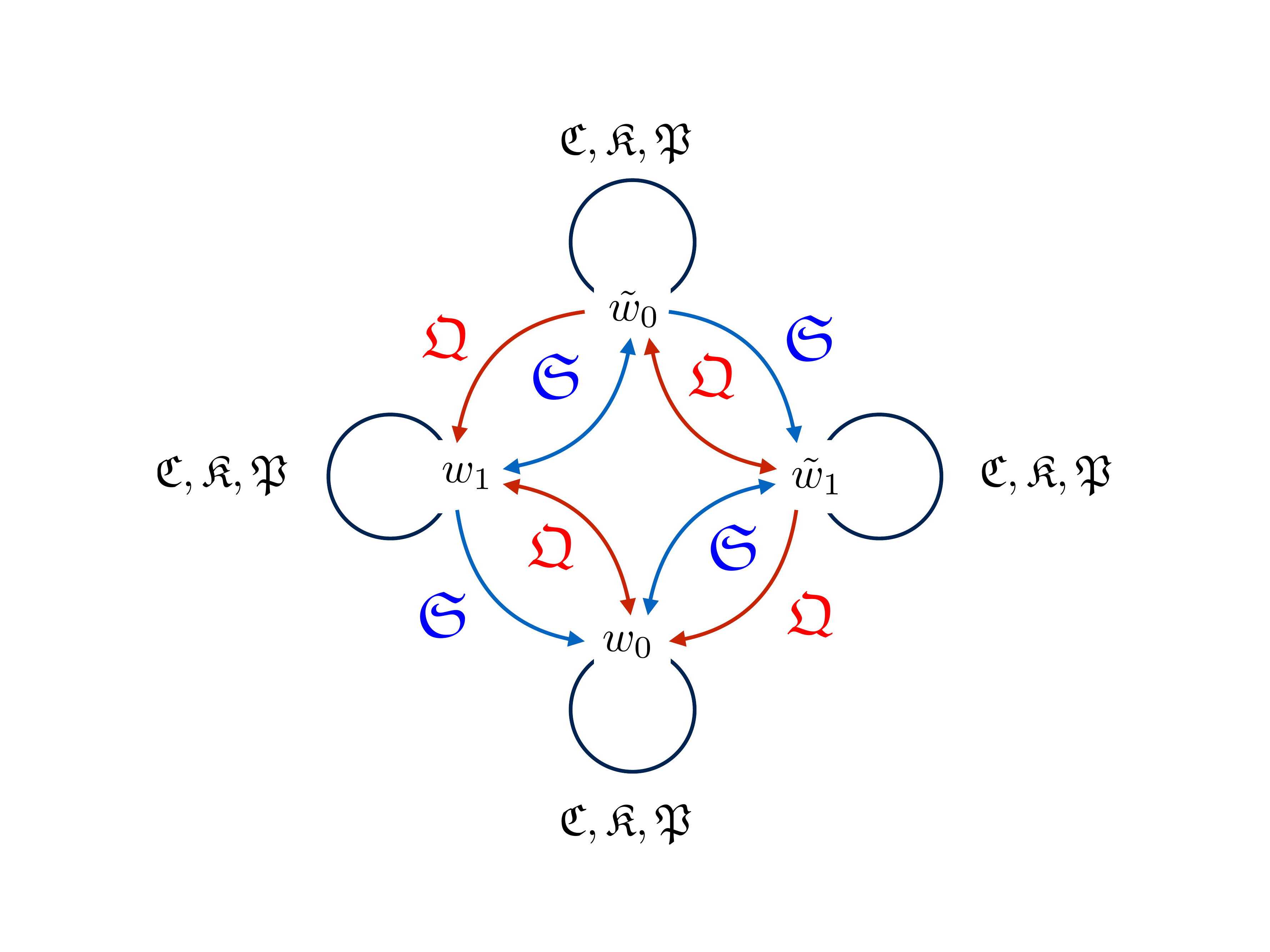}}
   \caption{The 4-dimensional module of the centrally-extended algebra.}
\label{fig:fig3}
\end{figure}

{\color{black} We have summarized the situation in figure \ref{fig:fig3}.} 

However, using the fact that
\begin{align}
& \alg{Q}\alg{S}|\tilde w_0\rangle = C |\tilde w_0\rangle + \frac {M}2|w_0\rangle \ ,
&& \alg{Q}\alg{S}|w_0\rangle = C|w_0\rangle + \frac {M}2|\tilde w_0\rangle \ ,
\\
& \alg{S}\alg{Q}|\tilde w_0\rangle = C |\tilde w_0\rangle - \frac {M}2|w_0\rangle \ ,
&& \alg{S}\alg{Q}|w_0\rangle = C|w_0\rangle - \frac{M}2|\tilde w_0\rangle \ ,
\end{align}
we see that defining the linear combinations
\begin{equation}
|\Phi_\pm\rangle = |w_0\rangle \pm |\tilde w_0\rangle \ ,
\end{equation}
implies
\begin{equation}
\alg{Q}\alg{S}|\Phi_\pm\rangle = \big(C+\frac{M}2\big)|\Phi_\pm\rangle \ , \qquad
\alg{S}\alg{Q}|\Phi_\pm\rangle = \big(C-\frac{M}2\big)|\Phi_\pm\rangle \ .
\end{equation}
Furthermore,
\begin{equation}
\alg{Q}|\Phi_\pm\rangle = \mp \frac{2C \mp M}{M}|w_1\rangle \pm \frac{2P}{M} |\tilde w_1\rangle\ , \qquad
\alg{S}|\Phi_\pm\rangle = \pm \frac{2C \pm M}{M}|\tilde w_1\rangle \mp \frac{2K}{M} |w_1\rangle\ .
\end{equation}
Using the definition of $M$ \eqref{mdef} one can easily see that
\begin{equation}
\alg{Q}|\Phi_\pm\rangle \propto \alg{S}|\Phi_\pm\rangle \propto |\Psi_\pm\rangle \ ,
\end{equation}
and hence the 4-dimensional representation we constructed is actually reducible
and is formed of two 2-dimensional representations
\begin{equation}
\{|\Phi_\pm\rangle,|\Psi_\pm\rangle\}\ .
\end{equation}

To conclude, let us briefly mention orthogonality. Here we will make use of the
real form of the algebra given in eq.~\eqref{hermi}, and the assumption that
$M$ is real. We then have
\begin{equation}
\langle\Phi_\mp|\Phi_\pm\rangle = \langle w_0|(\id + \frac1{M} ([\alg{Q},\alg{S}] - [\alg{Q},\alg{S}]^\dagger) - \frac 1{M^2} [\alg{Q},\alg{S}]^\dagger[\alg{Q},\alg{S}]|w_0\rangle\ .
\end{equation}
Using the conjugation relations we find that $[\alg{Q},\alg{S}]^\dagger =
[\alg{Q},\alg{S}]$. Furthermore, as $[\alg{Q},\alg{S}] =2 \alg{C} -2
\alg{S}\alg{Q} = -2\alg{C} +2\alg{Q}\alg{S}$ we find
\begin{equation}\begin{split}
\langle\Phi_\mp|\Phi_\pm\rangle &
= \langle w_0|\id + \frac 1{M^2} (2\alg{C}-2\alg{S}\alg{Q})(2\alg{C}-2\alg{Q}\alg{S})|w_0\rangle
= \langle w_0|\id - \frac 4{M^2}(\alg{C}^2 - \alg{P}\alg{K})|w_0\rangle
\\
& = (1- \frac{4(C^2-PK)}{M^2})\langle w_0|w_0\rangle = 0 \ .
\end{split}\end{equation}
Therefore, the two representations are orthogonal.

\corr{This construction can then be straightforwardly applied to the
4-dimensional representation arising as the tensor product of two of the
2-dimensional irreps of section \ref{massi}. Explicit details of this
construction are given in appendix \ref{appb} and will be particularly relevant
for the scattering theory discussed in section \ref{sec2}.  In particular, it
implies that the S-matrix for the scattering of two of the 2-dimensional
irreps is not completely fixed by symmetries up to an overall factor.}

Let us finally make the important observation that the arguments of this
section cannot be applied for the $M=0$ case (such as, for instance, the
scattering of two massless particles with the momenta taken at the bound-state
point\,\footnote{\label{footbsp}Here by {\it bound-state point} we simply mean
the value of momenta such that $\Delta^2(C) - \Delta(P) \Delta(K) = (m_1 +
m_2)^2 = 0$, namely $\ell_{ac}=0$ or $\ell_{bd}=0$ \corr{(see appendix \ref{appb}
for details)}. In fact, it is not clear if
there is a meaning of bound states for massless scattering
\cite{Zamolodchikov:1994za}.\fnsv}). In this case what we find is the analog of
the projective indecomposable representation of section \ref{gl1}. In
particular, one can check that, at $M=0$, the state $|\tilde{w}_0^{(0)}\rangle
\equiv [\alg{Q},\alg{S}]|w_0\rangle$ is such that
\begin{eqnarray}
\alg{Q}\alg{S}\, |\tilde{w}_0^{(0)}\rangle \,
= \, \alg{S}\alg{Q} \, |\tilde{w}_0^{(0)}\rangle \,
= \, C \, |\tilde{w}_0^{(0)}\rangle, \qquad \alg{Q} \, |\tilde{w}_0^{(0)}\rangle \, \propto \, \alg{S} \, |\tilde{w}_0^{(0)}\rangle,
\end{eqnarray}
where we have used $M^2=4(C^2 - PK) = 0$ to derive the last proportionality
statement. However, this is the only state which satisfies these properties,
meaning we do not have two solutions to these conditions (as we did in the $M
\neq 0$ case above). Therefore, there is only one irreducible 2-dimensional
block, containing the states $\{|\tilde{w}_0^{(0)}\rangle, \alg{Q}
|\tilde{w}_0^{(0)}\rangle\}$, and the 4-dimensional representation is reducible but
not fully reducible ({\em i.e.} it is indecomposable).

\section{S-matrix for massive modes of \texorpdfstring{$\mathbf{AdS_2 \times S^2}$}{AdS2 x S2}}\label{sec2}

In this section we study the S-matrix for the massive modes of the light-cone
gauge $AdS_2 \times S^2 \times T^6$ superstring. As mentioned in section
\ref{sec1} from the structure of the symmetry algebra and the integrability of
the theory we expect the S-matrix for the massive fields $y$, $z$, $\zeta$ and
$\chi$ to be constructed from the graded tensor product of two copies of an
S-matrix describing the scattering of $1+1$ massive modes, $\phi$ and $\psi$.
The former are defined in terms of the latter in \eqref{bhtens}.

The excitations $\phi$ and $\psi$ should transform in the massive representation
of $\mathfrak{psu}(1|1)\ltimes \mathbb{R}^3$ discussed in section \ref{massi}.
Their S-matrix is then fixed by demanding invariance under this symmetry
\begin{equation}
\Delta^{op}(\alg{J}) \mathbb{S} = \mathbb{S}\Delta(\alg{J})\ .
\end{equation}
Accounting for conservation of
\corr{the value of $(-1)^F$, where $F$ is the fermion number}, 
the most general form for the S-matrix is
\begin{eqnarray}
\label{redu}
&&\mathbb{S}\left|\phi\phi'\right> = S_1 \left|\phi\phi'\right> + Q_1 \left|\psi\psi'\right> \ , \qquad
\mathbb{S}\left|\psi\psi'\right> = S_2 \left|\psi\psi'\right> + Q_2 \left|\phi\phi'\right> \ ,\nonumber
\\
&&\mathbb{S}\left|\phi\psi'\right> = T_1 \left|\phi\psi'\right> + R_1 \left|\psi\phi'\right> \ , \qquad
\label{ans}
\mathbb{S}\left|\psi\phi'\right> = T_2 \left|\psi\phi'\right> + R_2 \left|\phi\psi'\right> \ ,
\end{eqnarray}
where $x^\pm$, $m$ are the kinematic variables associated to the
first particle and $x'^\pm$, $m'$ to the second particle, that is
\begin{equation}\label{yyss}
x^+ +\frac{1}{x^+} -x^{-}-\frac{1}{x^-} = \frac{2im}{\hh}\ , \qquad\quad
x'^+ +\frac{1}{x'^+} -x'^{-}-\frac{1}{x'^-} = \frac{2im'}{\hh}\ .
\end{equation}
As a consequence of the discussion in section \ref{32} this symmetry will only
fix the S-matrix up to two arbitrary functions. One of these functions can be
found by requiring the S-matrix also satisfies the Yang-Baxter equation along
with additional physical requirements. There are four solutions to the
Yang-Baxter equation, two of which we ignore as they violate crossing symmetry.
The other two are related by a sign. To fix the sign, we demand that in the BMN
limit (for details see section \ref{secpert}) the S-matrix reduces to the
identity operator.  The functions parametrizing the exact S-matrix \eqref{ans}
are then given by\,\footnote{Note that here we are choosing the branch so that
$\big(\frac{x^-}{x^+}\big)^{\#} = \big(\frac{x^+}{x^-}\big)^{-\#}$ for
$\#=\frac12,\,\frac14$ and similarly for $x'^\pm$. For $\pp\in[-\pi,\pi]$ this
corresponds to taking the branch cut on the negative real
axis.\fnsv}

\begin{align}\no
& S_1 = \sqrt{\frac{x^+\xpr^-}{x^-\xpr^+}}\frac{x^- - \xpr^+}{x^+ - \xpr^-} \frac{1+s_1}{2} \tPo \ ,
&& S_2 = \frac{1+s_2}{2} \tPo \ ,
\\\no
& T_1 =\sqrt{\frac{\xpr^-}{\xpr^+}} \frac{x^+ - \xpr^+}{x^+ - \xpr^-} \frac{1+t_1}{2}\tPo \ ,
&& T_2 =\sqrt{\frac{x^+}{x^-}} \frac{x^- - \xpr^-}{x^+ - \xpr^-}\frac{1+t_2}{2} \tPo \ ,
\\
& \frac{Q_1}{\a\a'} = \a\a'\, Q_2 = - \frac{i}{2} \,\sqrt[4]{\frac{x^-\xpr^+}{x^+\xpr^-}}
\frac{\h\h'}{x^+-\xpr^-} \frac f{x^-x'^+}\tPo \ , &&
\frac{\a'}{\a} R_1 = \frac{\a}{\a'}R_2 = -\frac{i}{2} \,
\sqrt[4]{\frac{x^+\xpr^-}{x^-\xpr^+}} \frac{\h\h'}{x^+ - \xpr^-} \tPo \ ,\label{exacta}
\end{align}
where
\begin{align}\label{exacta1}
f = \frac{\sqrt{\frac{x^+}{x^-}}(x^--\frac{1}{x^+}) - \sqrt{\frac{x'^+}{x'^-}}(x'^--\frac{1}{x'^+})}{1 - \frac{1}{x^+x^-x'^+x'^-}} \ , \qquad
s_1 = & \frac{1- \frac{1}{x^+\xpr^-}}{x^- - \xpr^+} f \ , \qquad s_2 = \frac{1-\frac{1}{x^-\xpr^+}}{x^+ - \xpr^-} f \ ,
\\
t_1 = & \frac{1 - \frac{1}{x^-\xpr^-}}{x^+ - \xpr^+} f \ , \qquad t_2 = \frac{1 - \frac{1}{x^+\xpr^+}}{x^- - \xpr^-} f \ .
\end{align}
$\tPo$ is an overall factor that sits outside the matrix structure and is not
fixed by symmetries or the Yang-Baxter equation. Let us emphasize that, as
discussed beneath eq.~\eqref{da}, when written in these variables the S-matrix
is independent of $m$ and $m'$, which can take any value. The limits $m\to0$
and $m'\to0$ are subtle however, and will be discussed in detail in section
\ref{massless}. Let us also note that if we take $\a$ to be given by
\eqref{nca}, which is the choice suitable for string theory, then $Q_1 = Q_2$
and $R_1 = R_2$. From now on we will take $\a$ to be given by this value.

The S-matrix \eqref{redu} can be thought of as a $4 \times 4$ block diagonal matrix
\begin{equation}
\left(\begin{array}{cccc} S_1 & Q_1 & 0 & 0 \\ Q_2 & S_2 & 0 & 0 \\
0& 0& T_1 & R_1 \\ 0&0& T_2 &R_2 \end{array}\right)\ .\end{equation}
One can then check that each of the two $2 \times 2$ blocks have equal trace
and determinant,
\begin{equation}\label{trdeteq}
S_1 + S_2 = T_1 + T_2 \ , \qquad S_1 S_2 - Q_1 Q_2 = T_1 T_2 - R_1 R_2 \ .
\end{equation}
The second of these equations is particularly important as it implies the
tensor product of two copies of the S-matrix possesses an additional $U(1)$
symmetry, which will be discussed further in section \ref{secu1} and appendix
\ref{appu1}.

\corr{
For completeness let us note that the two solutions that violate crossing symmetry are
given by $f=0$ and $f\to\infty$ (for the latter one should first
rescale $\tPo$ by $f^{-1}$ and then take $f \to \infty$). As $\phi$ and $\psi$
are real \cortwo{and 
the charge conjugation matrix diagonal, which will be demonstrated in the next section (cf. \eqref{eq:C})}, the two processes
\begin{equation}
\phi\,\phi \to \psi\,\psi \qquad \text{and} \qquad \phi\,\psi \to \psi\,\phi\ ,
\end{equation}
should be related by a crossing transformation. However, if $f$ vanishes then
so does the amplitude for the first of these processes, but not for the second.
Similarly, if $f\to\infty$ then the amplitude for the second process vanishes,
but not for the first. Consequently, in both cases the two processes cannot be
related by a crossing transformation and hence there is a violation of crossing
symmetry as claimed. 

It is interesting to note that taking $f=0$ and $f\to
\infty$ we recover the massive S-matrices of the $AdS_3 \times S^3 \times T^4$
light-cone gauge superstring \cite{Borsato:2013qpa,HT}. The symmetry is
enhanced accordingly from $\mathfrak{psu}(1|1)\ltimes \mathbb{R}^3$ to
$[\mathfrak{u}(1)\inplus \mathfrak{psu}(1|1)^2] \ltimes\mathfrak{u}(1)
\ltimes\mathbb{R}^3$. For the $AdS_3 \times S^3 \times T^4$ light-cone
gauge-fixed theory there is no issue with crossing symmetry as the fields are
complex. Therefore, the individual S-matrices do not map to themselves under
the crossing transformation, rather to a different S-matrix with the crossed
particle replaced by its antiparticle. Finally let us also point out that the
S-matrix relevant for the $AdS_2 \times S^2 \times T^6$ light-cone gauge
superstring, see eqs.~\eqref{redu} and \eqref{exacta}, is a linear combination,
with coefficients depending on $x^\pm$ and $x'^\pm$, of the $f=0$ and $f\to
\infty$ S-matrices. It is non-trivial that such a combination exists with
unitarity, crossing symmetry and the Yang-Baxter equation all satisfied.}
\subsection{The overall factor and crossing symmetry}

As currently written the factor $\tPo$ is neither a phase factor or
antisymmetric. Indeed, given the reality conditions $(x^\pm)^* = x^\mp$ and
$(x'^\pm)^* = x^\mp$, the functions $f$, $s_{1,2}$ and $t_{1,2}$ satisfy the
following relations:
\begin{alignat}{3}
f^* & = f \ , \qquad & s_{1,2}^* & = s^{\vphantom{*}}_{2,1} \ , \qquad & t_{1,2}^* &= t^{\vphantom{*}}_{2,1}\ ,
\\f(x',x) & = - f(x',x) \ , \qquad & s_{1,2}(x',x) & = s_{2,1}(x,x') \ , \qquad & t_{1,2}(x',x) &= t_{1,2}(x,x') \ .
\end{alignat}
{\color{black} Notice that, if we consider the $m=m'$ case, then
on-shell ({\it i.e.} when the dispersion relations \eqref{yyss} are satisfied) we
have $t_1 \approx t_2$. Given the reality conditions, this means in particular that $t_1, t_2$ are real.\fnsv

Based on this, and as a consequence of braiding and QFT unitarity,} the overall factor
should satisfy\,\footnote{Note that the $AdS_5 \times S^5$ S-matrix contains
copies of the $2 \times 2$ block:
\begin{equation*}
\bigg(\begin{array}{cc}
2T_1 & 2R_1 \\ 2R_2 & 2T_2 \end{array}\bigg) \bigg|_{t_i = 0} \ .
\end{equation*}
Taking into account the factor of 2, when $t_i=0$ \eqref{unit} simplifies to
$\tPo\tPo^* = \tPo(x,\xpr)\tPo(\xpr,x)=1$ so that $\tPo$ is an antisymmetric
phase factor. This is the familiar $AdS_5 \times S^5$ story.\fnsv}
\begin{equation}\label{unit}
\tPo\tPo^* = \tPo(x,\xpr) \tPo(\xpr,x)
= \frac{4(x^- - \xpr^+)(x^+ - \xpr^-)}{(x^+ - \xpr^+)(x^- - \xpr^-) (1+t_1)(1+t_2) - (x^+ - x^-)(\xpr^+ -\xpr^-)}
\corr{\equiv N(x,x')}\ . 
\end{equation}
\corr{To isolate an antisymmetric phase factor, we can define $\Po$ as follows:
\begin{equation}
\Po =
 \det \bigg( \begin{array}{cc}
S_1 & Q_1 \\ Q_2 & S_2 \end{array}\bigg) =
 \det \bigg( \begin{array}{cc}
T_1 & R_1 \\ R_2 & T_2 \end{array}\bigg) \ {\color{black} \equiv \, \exp{\corp{i}\theta(x,x')}},
\end{equation}
{\color{black} where $\theta(x,x')$ is an antisymmetric {\it phase shift}, \corp{{\em i.e.}} $\theta(x,y) = - \theta(y,x)$, and} the second equality follows from eq.~\eqref{trdeteq}. {\color{black} We then have}
that $\Po$ is proportional to $\tPo^2$, and hence is a natural phase to consider
recalling that
the full S-matrix for the massive modes is given by the tensor product
of two of the factor S-matrices \eqref{redu}.}
As claimed the unitarity
conditions for $\Po$ are then
\begin{equation}\label{unit2}
\Po\Po{}^{ \! *} = \Po(x,\xpr) \Po(\xpr,x) = 1\ .
\end{equation}

\

\def\arg{\corr{(x,x')}} 
\def\argc{\corr{(x',\bar x)}}
\def\argca{\corr{(\bar x,x')}}

Crossing symmetry provides an additional constraint on the overall factor
$\tPo$, which takes the form
\begin{equation}\label{cross}
\tPo\argc = s_2\arg \, \tPo\arg\ , 
\end{equation}
where the
``crossed'' Zhukovsky variables $\bar x^\pm$ are, as usual, given by
\begin{equation}
\bar x^\pm = \frac1{x^\pm}\ ,
\end{equation}
corresponding to $\bar e = -e$ and $\bar \pp = -\pp$. It is useful to note
that we have the following identities
\begin{equation}
s_{1,2}\argc = s_{1,2}^{-1}\arg \ , \qquad t_{1,2}\argc = t_{2,1}^{-1}\arg \ . 
\end{equation}
\corr{Using the braiding unitarity relation \eqref{unit} it is simple 
to recast \eqref{cross} in the more familiar form
\begin{equation}\label{crossa}
\tPo\arg\tPo\argca = \frac{N(\bar x,x')}{s_2(x,x')}\ .
\end{equation}
This relation then translates to the following rather complicated
constraint for the antisymmetric phase factor $\Po$ 
{\color{black}
\begin{equation}\label{cross2}
\Po\arg\Po\argca = \frac{S_1 S_2 - Q_1 Q_2}{S_1 S_2 + R_1 R_2}
= \frac{T_1 T_2 - R_1 R_2}{T_1 T_2+Q_1 Q_2}\equiv f_2\arg\ ,
\end{equation}
}
and hence it appears that we either have a simple crossing relation or simple
unitarity relations.}

Using Hopf algebra arguments, we have checked that crossing symmetry is present
for the representation of interest for any value of $m$ and $m'$. Denoting the
symmetry algebra as $\cal{A}$, the antipode $\Sigma$ is found from the defining
rule
\begin{equation}\label{rule}
\mu \, (\Sigma \otimes\id) \, \Delta = \eta \, \epsilon\ ,
\end{equation}
where $\mu$ is the multiplication map, $\eta : \mathbb{C} \to {\cal{A}}$ is the
unit and $\epsilon : {\cal{A}} \to \mathbb{C}$ is the counit, which annihilates
all generators apart from $\id$ and $e^{i\pp}$ (acting on which, it
returns $1$). The antipode being a Lie algebra anti-homomorphism, we simply
need to derive
\begin{equation}
\Sigma (\gen{Q}) = - e^{- i \frac{\pp}{2}}\gen{Q}, \qquad \Sigma (\gen{Q}) = - e^{i \frac{\pp}{2}}\gen{G}, \qquad \Sigma(\id) = \id, \qquad \Sigma(e^{i\pp}) = e^{-i\pp}\ .
\end{equation}
This map is idempotent and therefore equal to its inverse. We impose
\begin{equation}\label{eq:underl}
\Sigma \big(\gen{J}(x^\pm)\big) = \mathscr{C}^{-1} \bigg[\gen{J}\bigg(\frac{1}{x^\pm}\bigg)\bigg]^{\cortwo{st}} \mathscr{C}\ ,
\end{equation}
where $\mathscr{C}$ is the charge conjugation matrix
\begin{equation}\label{eq:C}
\mathscr{C} = \begin{pmatrix}1&0\\0&i\end{pmatrix}\ ,
\end{equation}
and the label ${}^{\cortwo{st}}$ denotes supertransposition. The fundamental crossing
relation for an \corr{abstract} R-matrix\,\footnote{For our purposes, S-matrices will be
representations of abstract R-matrices.\fnsv} is then given by ({\em cf.}
\cite{Janik:2006dc})
\begin{equation}
(\Sigma \otimes \id) R = R^{-1} = (\id \otimes \Sigma^{-1}) R\ ,
\end{equation}
\corr{ which projects into the representation of interest as
\begin{equation}\label{matricial}
(\mathscr{C}^{-1} \otimes \id) \mathbb{S}^{\cortwo{st}_1}(\bar x, x') (\mathscr{C} \otimes \id) \mathbb{S} (x, x')
= \id \otimes \id\ ,
\end{equation}
and an analogous equation for the second factor. Here ${}^{\cortwo{st}_i}$ denotes the
supertranspose for factor $i$, {\color{black} and we are using the Hopf algebra convention for the S-matrix crossing \cite{Janik:2006dc} (see \cite{rev1} for the convention used in the field theory literature)}. The S-matrix \eqref{ans} with parametrizing
functions \eqref{exacta} satisfies this relation provided the overall factor
satisfies the crossing equation given in \eqref{cross}.}

\corr{It is important to note that the crossing equations given above are
somewhat formal as we have not specified a path on the rapidity plane. To
specify such a path we would need to know the precise form of the dispersion
relation, and hence its uniformization. {\color{black} In particular, there is still the logical possibility that $m$ and $m'$ are themselves momentum-dependent functions (which \corp{should be invariant under crossing}). 
This possibility would not alter the analysis we have performed so far.}
  In the scenario that $m$ and $m'$ are
non-vanishing and constant the dispersion relation becomes the same as in the
$AdS_5 \times S^5$ light-cone gauge string theory and the analytic continuation
should be the same as in that case {\color{black} \cite{Janik:2006dc,Arutyunov:2007tc,Arutyunov:2009kf}}.} 

\corr{In spite of our lack of knowledge of the {\color{black} complete} dispersion relation, one
thing we can investigate is double crossing \cite{Janik:2006dc}.\footnote{We would like to
thank the referee for \corp{suggesting the consideration of double crossing.}} In particular, the
left-hand sides of \eqref{crossa} and \eqref{cross2} are symmetric under $x
\leftrightarrow \bar x$, however the right-hand sides are not. This asymmetry
encodes the fact that the overall factor should not be a meromorphic function of the
parameters that uniformize the dispersion relation (generalized rapidities).
Furthermore, as a consistency check, one can \corp{confirm} 
that the following equality holds true:
{\color{black}
\begin{equation}\label{ratio}
\corp{\Omega(x,x') \, \equiv \, \frac{\tPo^2 (x,x')}{\tPo^2 (\bar {\bar x},x')} \ . \ \frac{\Po (\bar {\bar x},x')}{\Po (x,x')}
= \, \left(\frac{N(\bar x,x')}{s_2(x,x')}\right)^2 \ . \ \left(\frac{s_2(\bar x,x')}{N(x,x')}\right)^2
\ . \ \frac{f_2(\bar x,x')}{f_2\arg} \, = \, 1 \ .}
\end{equation}
This is obtained by comparing the ratio of the right-hand side of \eqref{crossa} (squared) to the same quantity with $x
\to \bar x$, against the corresponding ratio for the right-hand side of
\eqref{cross2}. The fact that $\Omega = 1$ confirms that $\tPo^2$ and $\Po$ differ only by a factor that behaves like
a rational function under double crossing, as is expected. 

It is easy to convince oneself that the ratio $\frac{f_2(\bar x,x')}{f_2\arg}$ encodes the
discontinuity of the overall S-matrix factor across branch cuts in the, as yet unknown,
rapidity plane. It is of interest to note that this ratio} 
differs
from the corresponding one in the $AdS_5 \times S^5$ case, suggesting that the
analytic structure of the $AdS_2 \times S^2$ light-cone gauge-fixed theories is not the same.
To understand crossing symmetry and the phase in more detail clearly
requires a deeper knowledge of the dispersion relation, which, as it is not {\color{black} entirely} fixed
by symmetries, we leave for future investigation.}

\subsection{Comparison with perturbation theory}\label{secpert}

Defining the effective string tension
\begin{equation}\label{ef}
\coup = \frac{R^2}{2\pi\a'} \ ,
\end{equation}
the tree-level S-matrix for the scattering of massive modes in the light-cone
gauge $AdS_2 \times S^2 \times T^6$ superstring following from near-BMN
perturbation theory can be be found by suitably truncating the corresponding
result for $AdS_5 \times S^5$ or $AdS_3 \times S^3 \times T^4$ \cite{pert}
(various components were also computed in \cite{amsw}). This gives
\begin{align}
S_1 = & \ 1 + \frac{i}{4\coup}\big[(1-2\ra)(e' p - e p') + l_1 \big] + \mathcal{O}(\frac1{\coup^2}) \ , \no
\\ S_2 = & \ 1 + \frac{i}{4\coup}\big[(1-2\ra)(e' p - e p') - l_1 \big] + \mathcal{O}(\frac1{\coup^2}) \ , \no
\\ T_1 = & \ 1 + \frac{i}{4\coup}\big[(1-2\ra)(e' p - e p') - l_2 \big] + \mathcal{O}(\frac1{\coup^2}) \ , \no
\\ T_2 = & \ 1 + \frac{i}{4\coup}\big[(1-2\ra)(e' p - e p') + l_2 \big] + \mathcal{O}(\frac1{\coup^2}) \ , \no
\\ Q_1 = & \ Q_2 = \frac{i}{2 \coup} l_3 + \mathcal{O}(\frac1{\coup^2}) \ , \qquad
R_1 = R_2 = - \frac{i}{2\coup} l_4 + \mathcal{O}(\frac1{\coup^2}) \ , \label{pertads3}
\end{align}
where the functions $l_i$ are defined as
\begin{align}
l_1(p,p') = & \frac{p^2+p'^2}{e'p-ep'} \ , \qquad \qquad
l_2(p,p') = \frac{p^2-p'^2}{e'p-ep'} \ , \no
\\
l_3(p,p') = & - \frac{pp'}{2(e'p-ep')}\big[\sqrt{(e+p)(e'-p')} - \sqrt{(e-p)(e'+p')}\big] \no \ ,
\\
l_4(p,p') = & - \frac{pp'}{2(e'p-ep')}\big[\sqrt{(e+p)(e'-p')} + \sqrt{(e-p)(e'+p')}\big] \no \ .
\end{align}
The parameter $\ra$ is the standard gauge-fixing parameter of the uniform
light-cone gauge \cite{lcg}. In \cite{amsw} it was shown that to one-loop the
near-BMN dispersion relation is given by
\begin{equation}\label{nearbmn}
e^2 = 1+ p^2 + \mathcal{O}(h^{-2})\ .
\end{equation}

The one-loop near-BMN result can be constructed via unitarity methods following
\cite{uc}. \corr{As expected from unitarity methods, this will certainly give
the correct logarithmic terms in the one-loop S-matrix, and indeed this has
already been argued in \cite{amsw}. However, the prescription given in
\cite{uc} is also conjectured to give the correct rational terms for integrable
theories. Under this assumption we find that} 
the \corr{one-loop} S-matrix takes the following form 
\begin{align}
S_1 = & \ \exp\big\{\frac{i}{4\coup}(1-2\ra)(e' p - e p')\big\} \ \s_{AdS_2} \ \big[1 + \frac{i}{4\coup}l_1 - \frac{\ell}{32\coup^2}\, \big] + \mathcal{O}(\frac1{\coup^3}) \ , \no
\\
S_2 = & \ \exp\big\{\frac{i}{4\coup}(1-2\ra)(e' p - e p')\big\} \ \s_{AdS_2} \ \big[1 - \frac{i}{4\coup}l_1 - \frac{\ell}{32\coup^2}\, \big] + \mathcal{O}(\frac1{\coup^3}) \ , \no
\\
T_1 = & \ \exp\big\{\frac{i}{4\coup}(1-2\ra)(e' p - e p')\big\} \ \s_{AdS_2} \ \big[1 - \frac{i}{4\coup}l_2 - \frac{\ell}{32\coup^2} \, \big] + \mathcal{O}(\frac1{\coup^3}) \ , \no
\\
T_2 = & \ \exp\big\{\frac{i}{4\coup}(1-2\ra)(e' p - e p')\big\} \ \s_{AdS_2} \ \big[1 + \frac{i}{4\coup}l_2 - \frac{\ell}{32\coup^2} \, \big] + \mathcal{O}(\frac1{\coup^3}) \ , \no
\\
Q_1 = & \ Q_2 = \exp\big\{\frac{i}{4\coup}(1-2\ra)(e' p - e p')\big\} \ \s_{AdS_2} \ \big[\frac{i}{2\coup}l_3 \, \big] + \mathcal{O}(\frac1{\coup^3}) \ , \no
\\
R_1 = & \ R_2 = \exp\big\{\frac{i}{4\coup}(1-2\ra)(e' p - e p')\big\} \ \s_{AdS_2} \ \big[-\frac{i}{2\coup}l_4 \, \big] + \mathcal{O}(\frac1{\coup^3}) \ , \label{polads2}
\end{align}
where the expansion of the phase factor $\s_{AdS_2}$ is given by
\begin{equation}\begin{split}
\sigma_{AdS_2} = \exp \big\{& \frac{i}{8\pi\,\coup^2}\frac{p^2p'{}^2\big((e'p-ep') - (ee'-pp')\operatorname{arsinh}[e'p-ep']\big)}{(e'p-ep')^2} + \mathcal{O}(\frac{1}{\coup^3}) \big\}\ ,
\end{split}\end{equation}
while
\begin{equation}
\ell =\frac{p^4 + p'{}^4 +2 p^2 p'{}^2 (e e' - p p')}{(e'p - ep')^2} \ ,
\end{equation}
is fixed by the requirement of unitarity. \corr{As observed in \cite{amsw} the one-loop
logarithms are consistent with the one-loop phase being related to the Hernandez-Lopez
phase \cite{hl}.} 

We define the near-BMN expansion of the exact result as follows
\begin{equation}\begin{split}\label{annn}
& \ee = e\ , \qquad
m = \rho_3 + \rho_4 h^{-1} + \mathcal{O}(h^{-2}) \ , \qquad
\pp = \frac{p}{h\big(\rho_5 + \rho_6 h^{-1} + \mathcal{O}(h^{-2})\big)}\ ,
\\
& \hh = h\big(\rho_1 + \rho_2 h^{-1} + \mathcal{O}(h^{-2})\big) \ ,
\end{split}\end{equation}
and similarly for $\ee',\pp'$ and $m'$. Here for generality we have allowed for
various rescalings, however, for simplicity we will assume that the $\rho_i$
are constants.\footnote{To be completely general, one could in principle let
$\ee,\,m$ and $\pp$ be arbitrary functions of $e$ and $p$. However, naively
truncating the classical/tree-level results for $AdS_5 \times S^5$ and $AdS_3
\times S^3 \times T^4$, for example \cite{Arutyunov:2006ak,BogdanLatest}, to
the massive sector of $AdS_2 \times S^2 \times T^6$ the ansatz \eqref{annn}
seems reasonable. Of course to check this claim one should construct the
light-cone gauge symmetry algebra
explicitly.\fnsv}

{\color{black} Let us remark that
in this paper we are considering the $AdS_2 \times S^2 \times T^6$
background supported by Ramond-Ramond fluxes \cite{Sorokin:2011rr}, and hence
the light-cone gauge-fixed theory should be parity invariant
\cite{Murugan:2012mf,amsw}. Therefore, if it were the case that $m$ receives
quantum corrections depending on the momentum they should respect \cortwo{the corresponding} constraint.\footnote{\label{fooot}\cortwo{In section \ref{massless} we will} study the $m \to 0$ limit as a massless regime, with the {\it proviso} that
\cortwo{if it were the case that $m$ becomes momentum-dependent at a quantum level, this limit would no longer be relevant for the massless modes of the superstring. 
This issue should be addressed through a more detailed study of the off-shell symmetry algebra of the theory and its representations \cite{Arutyunov:2006ak,BogdanLatest}.}\fnsv}
This is in contrast to backgrounds partially (or wholly) supported by Neveu-Schwarz
flux, for which $m$ may have a dependence on $p$ that breaks parity (see, for
example, \cite{dr} for discussions of the dispersion relation of the $AdS_3
\times S^3 \times T^4$ light-cone gauge-fixed theory supported by a mix of
fluxes).  
It is worth noting that the $AdS_2 \times S^2 \times T^6$ background
can also be supported by a mixture of Ramond-Ramond and Neveu-Schwarz fluxes
\cite{sym} and it would be interesting to see how the presence of the latter
affects the representations discussed in this paper. }

 Expanding the exact
dispersion relation \eqref{apiii} in the near-BMN regime, we recover
\eqref{nearbmn} if we take
\begin{equation}
\rho_5=\rho_1 \ , \qquad \rho_6 = \rho_2 \ , \qquad
\rho_3 = 1\ , \qquad \rho_4=0 \ .
\end{equation}
Further expanding the exact S-matrix \eqref{exacta} in the near-BMN regime,
taking $\a$ given by \eqref{nca}, and fixing the overall factor $\tPo$ such that
any one of the eight amplitudes agrees with perturbation theory, we find that,
so long as
\begin{equation}
\rho_1 = 1 \ ,
\end{equation}
the remaining seven also agree with perturbation theory, \eqref{pertads3} and
\eqref{polads2}.

\section{Yangian symmetry}\label{Yangian}

\subsection{Massive case}

In this section we would like to discuss the issue of Yangian symmetry. The
first observation is that, in the massive case (we can fix $m=m'=1$ for the
purposes of this section), we could not apply the same standard Yangian
symmetry of the R-matrix which works for the massless case (see section
\ref{masslessyang}). The massive representation is a long one ({\em cf.}
section \ref{massi}), and a similar result was found for long representations
of $\mathfrak{psu}(2|2)\ltimes \mathbb{R}^3$ \cite{Arutyunov:2009pw}. The long
representations studied in \cite{Arutyunov:2009pw} bear a strong resemblance to
the ones in this paper, up to the different dimensionality.

We proceed by postulating the commutation relations of
the standard $\alg{sl}(1|1)$ Yangian in Drinfeld's second realization
\cite{Drinfeld:1987sy,Spill:2008tp} (with central extensions)
\begin{equation} \label{eq:Lie}
\begin{gathered}
\acomm{\gen{e}_m}{\gen{f}_n} = -\gen{h}_{m+n}\ , \qquad
\acomm{\gen{e}_m}{\gen{e}_n} = \gen{p}_{m+n}\ , \qquad
\acomm{\gen{f}_m}{\gen{f}_n} = \gen{p}_{m+n}^\dagger\ ,
\qquad \comm{\gen{h}_m}{\cdot} = \comm{\gen{p}_m}{\cdot}=\comm{\gen{p}_m^\dagger}{\cdot}=0\ . \\
\end{gathered}
\end{equation}
One can check that the coproducts obtained from
\begin{equation}
\begin{aligned}\label{Yangcopr}
\Delta({\gen{e}}_1) &= \gen{e}_1 \otimes \id + e^{i \frac{\pp}{2}} \otimes \gen{e}_1 + \gen{h}_0 \, e^{i \frac{\pp}{2}} \otimes \gen{e}_0\ , \\
\Delta(\gen{f}_1) &= \gen{f}_1 \otimes \id + e^{-i \frac{\pp}{2}} \otimes \gen{f}_1 + \gen{f}_0 \otimes \gen{h}_0\ ,
\end{aligned}
\end{equation}
and their opposites satisfy the defining relations \eqref{eq:Lie} and hence
provide homomorphisms of the Yangian. The antipode $\Sigma$ can be easily found
from \eqref{Yangcopr} using the defining property
\begin{equation}\label{rule2}
\mu \, (\Sigma \otimes \id) \, \Delta = \eta \, \epsilon\ ,
\end{equation}
where $\epsilon$ annihilates all level $1$ generators. Combined, this defines
the Hopf algebra structure of the standard Yangian.

One can construct a family of representations of the Yangian \eqref{Yangcopr}
starting from a slightly simpler level-zero (Lie algebra) representation
compared to the one we use in section \ref{massi}. Determining the level $1$
generators in this representation, we can obtain all the central elements up to
and including level $2$, together with their coproducts and opposite
coproducts.\footnote{In the absence of non-central Cartan elements, we cannot
mechanically generate the level $2$ and higher supercharges and they would have
to be guessed. However we do not need them for the sake of this argument.\fnsv}
Following the strategy of \cite{Arutyunov:2009pw}, one can check whether all
the central coproducts are co-commutative, as this is a necessary condition for
the existence of an R-matrix scattering two such representations (see footnote
\ref{foo10}). We found that
\begin{eqnarray}
\Delta^{op}(\gen{p}_2) \neq \Delta(\gen{p}_2)\ ,
\end{eqnarray}
for all members of the family of representations. This implies that at least
one representation of the standard Yangian does not admit an R-matrix,
excluding the existence of a universal R-matrix.

However, it is likely that the massive R-matrix may admit a coproduct which is
not precisely the same as for massless representations, but still of the type
found in \cite{Beisert:2007ds}. Moreover, considerations as in footnote 3 of
\cite{Arutyunov:2009pw} are likely to apply. We leave this investigation
for future work.

\subsection{Massless case}\label{masslessyang}

The situation is different for the massless limit $m=m'=0$ {\color{black} (see the discussion at the beginning of section \ref{massless})}. In this case, in
the absence of the central extensions ($b=c=0$, {\em i.e.} considering again
the $\mathfrak{gl}(1|1)$ algebra), the representation would become one of the
reducible but indecomposable modules of section \ref{gl1}. In fact, in that
case the condition $m = ad - bc = ad = 0$ would force one of the fermionic
generators to be identically zero. The indecomposable would then be made up of
{\it short} 1-dimensional $\mathfrak{gl}(1|1)$ irreps. This suggests that the
Yangian might now be straightforwardly derived from the standard one.

{\color{black} The fact that $m=0$ effectively works as a shortening condition, and the consequence 
that \corp{this allows for} 
the existence of a Yangian representation, gives us significant encouragement that 
$m=0$ might be protected against quantum corrections in the full theory. 
This is also corroborated by explicit perturbative results, 
which have not yet found \corp{any} evidence for \corp{a} quantum lift of this condition 
\corp{(see, for instance, \cite{Sorokin:2011rr,Murugan:2012mf})}. 
Moreover, \corp{the subgroup of $SO(6)$ controlling the symmetry of the massless
sector might allow one to construct a mechanism protecting the $m=0$ condition, analogous to the one described in \cite{BogdanLatest}
for $AdS_3 \times S^3 \times T^4$.}}

Indeed, this time we construct an evaluation representation of the Yangian
\eqref{Yangcopr}
\begin{eqnarray}
\gen{e}_1 = u \, \gen{e}_0 \, = \, u \, \mathfrak{Q}\ , \qquad
\gen{f}_1 = u \, \gen{f}_0 \, = \, u \, \mathfrak{G}\ , \qquad
u = \frac{i \hh}{x^-}\ ,
\end{eqnarray}
starting from the level $0$ one we consider in section \ref{massi},
specializing to $m=0$. Due to the additional parameters compared to the
$\mathfrak{gl}(1|1)$ case, the representation remains generically irreducible.
Nevertheless, the obstruction encountered in the massive case is no longer
present, {\em i.e.} all central charges we can build are co-commutative and in
fact the R-matrix (for $m=m'=0$) can be shown to be invariant under the
standard Yangian. This is reminiscent of the $AdS_5 \times S^5$ case, where the
Yangian for short representations does not directly transfer to long ones as it
stands \cite{Beisert:2007ds,Torrielli:2011gg}.

The crossing symmetry transformation reveals an interesting property, related
to what was observed in \cite{Borsato:2013qpa} for the case of $AdS_3 \times
S^3 \times T^4$, namely the existence of two different Yangian spectral
(evaluation) parameters for the particle and the anti-particle representations.
Here, the difference is superficial, as the massless condition makes the two
spectral parameters coincide. In fact,
the antipode obtained from applying \eqref{rule2} reads
\begin{eqnarray}
\label{podo}
\Sigma (\gen{e}_1) = - e^{-i \frac{\pp}{2}} \, (\gen{e}_1 + \gen{e}_0 \, \gen{h}_0)\ , \qquad
\Sigma (\gen{f}_1) = - e^{i \frac{\pp}{2}} \, (\gen{f}_1 + \gen{f}_0 \, \gen{h}_0)\ .
\end{eqnarray}
This effectively amounts to a shift in the spectral parameter $u$ by one of the
central elements. When plugging this into the relation
\begin{equation}
\label{crossato}
\Sigma \big({\gen{j}_1}(x^\pm)\big) = \mathscr{C}^{-1} \bigg[{\gen{j^a}_1}\bigg(\frac{1}{x^\pm}\bigg)\bigg]^{\cortwo{st}} \mathscr{C}\ ,
\end{equation}
and postulating that the anti-particle representation is also of evaluation
type, that is
\begin{eqnarray}
\gen{e^a}_1 = u_a \, \mathfrak{Q}\ , \qquad \gen{f^a}_1 = u_a \, \mathfrak{G} \ ,
\end{eqnarray}
we see that the conditions \eqref{crossato} and \eqref{podo} reduce to the same
equation that holds true for the level $0$ charges, {\em i.e.}
\eqref{eq:underl}, provided that the anti-particle spectral parameter is chosen
to be
\begin{equation}
u_a = i {\hh}\, x^+\ .
\end{equation}
For massless particles,
\begin{eqnarray}
u=u_a\ .
\end{eqnarray}

\section{S-matrix for massless modes}\label{massless}

\corr{In this section we investigate the $m \to 0$ and $m' \to 0$ limits of the S-matrix 
constructed in section \ref{sec2}. From the dispersion relation \eqref{apiii}
{\color{black} and under the assumption of constant $m$ and $m'$, we may consider it natural to interpret these} as massless limits.

While in principle {\color{black} these limits are already of interest in their own right}, given the 
Yangian symmetry discussed in section \ref{masslessyang}, the resulting
S-matrices may also be relevant for the scattering of massless modes in the
$AdS_2 \times S^2 \times T^6$ light-cone gauge string theory.  Indeed, for
the $AdS_3 \times S^3 \times T^4$ light-cone gauge-fixed theory the massless
modes transformed in the same {\color{black} type of} representations as the massive modes (with
vanishing mass and up to a {\color{black} suitable identification} of highest weight states)
\cite{BogdanLatest}. Motivated by this, one may conjecture that the S-matrices
constructed below can be used to build the S-matrices describing scattering
processes involving massless modes (under the assumption that they remain
massless \cortwo{and $m$ and $m'$ remain zero} at a quantum level \cortwo{-- see {\color{black} the discussion below \eqref{annn} and footnote \ref{fooot}}})
in the $AdS_2 \times S^2 \times T^6$ light-cone
gauge superstring.}

\subsection{Derivation from Yangian invariance}\label{maya}

The S-matrix describing the scattering of two massless excitations can be
directly obtained by imposing Lie algebra and Yangian invariance for two $m=0$
representations of section \ref{massi}, or as an $m,m' \to 0$ limit of the
massive S-matrix. In the latter case, one has to treat various $\frac{0}{0}$
limiting expressions, which come from the function $f$ in
eq.~\eqref{exacta1}.\footnote{This is somehow reminiscent of the relativistic
case \cite{Zamol2}.\fnsv} Taking care when resolving these singular limits we
find agreement with the result from imposing Yangian invariance. In the
massless limit the dispersion relation in terms of the Zhukovsky variables
takes the form \cite{BogdanLatest}\footnote{There is a second solution
$x^+=x^-$, however, this corresponds to $\pp=0$ and therefore is not physically
sensible.\fnsv}
\begin{equation}
x^+ = \frac{1}{x^-}\ .
\end{equation}
In terms of the energy and momenta this translates to
\begin{equation}
\ee^2= 4\hh^2\sin^2\frac{\pp}{2}\qquad \Rightarrow \qquad
\ee = 2 \hh\,\big|\sin\frac{\pp}2\big| \ ,
\end{equation}
and hence there are two branches of the dispersion relation depending on the
sign of $\sin\frac{\pp}{2}$
\cite{BogdanLatest}
\begin{align}
x^+ = \, & \sigma e^{i \, \frac{\pp}2}\ , \quad
x^- = \frac{1}{x^+}\ , \quad\s=\pm1 \ ,\qquad \qquad
x'^+ =\,\sigma' e^{i \, \frac{\pp'}2}\ , \quad
x'^- = \frac{1}{x'^+}\ , \quad \s'=\pm1\ .
\end{align}
In the following we will use the convention that $\s=+1$ corresponds to a
particle moving from left spatial infinity to right spatial infinity, {\em
i.e.} right-moving, while $\s=-1$ corresponds to a left-moving particle.

\corr{
Although the doubly-branched
dispersion relation $\ee = 2\hh |\sin \frac{\pp}{2}|$ is non-relativistic,
there are some similarities with the kinematics of massless relativistic
scattering. Following \cite{Zamol2}, in the relativistic case one
has
\begin{eqnarray}
e = \, \frac{m_0}{2} \, e^u\ , \qquad p = \, \pm \frac{m_0}{2} \, e^u, \qquad m_0 , \, u \in \mathbb{R} \ .
\end{eqnarray}
A boost sends the rapidity $u \to u + \lambda$, with $\lambda \in \mathbb{R}$,
hence the two branches can never be connected by such a transformation. In the
non-relativistic case we have the two branches
\begin{align}
\nonumber
&\frac{i \, \ee}{\hh} \,= \, \Big[x^+ - \frac{1}{x^+}\Big]\ , \qquad
\pp = - 2 i \log x^+ \in [0,\pi] \ ,
\\
&\frac{i \, \ee}{\hh} \,= \, \Big[x^+ - \frac{1}{x^+}\Big]\ , \qquad\label{mome}
\pp = - 2 i \log (- x^+) \in [-\pi,0]\ ,
\end{align}
with $x^+$ a pure phase for real momentum and energy. As the S-matrix is not of
difference form there is a priori no notion of boosts and hence it is not clear
if the presence of two branches represents an obstruction to interpreting the
$\s=\s'=\pm1$ scattering. However, as pointed out in \cite{BogdanLatest}, while
the small momentum dispersion relation is relativistic, for the exact
non-relativistic dispersion relation, the group velocity $\vv = \frac{\partial
\ee}{\partial \pp}$ is a non-trivial function of $\pp$ and hence one may hope
to give a physical interpretation to the $\s=\s'=\pm1$ scattering.}

\

For $\s=\s'=+1$, the Yangian invariance fixes the S-matrix up to {\it two}
undetermined functions $\chi_{1,2}^{++}$:
\begin{align}\nonumber
& S_1 = -S_2 = \frac{1}{\sin \frac{1}{4}(\pp + \pp')} \Bigg[\chi_1^{++} \, \sin \frac{1}{4}(\pp - \pp') + \, \chi_2^{++} \, \sqrt{\sin \frac{\pp\vphantom{\pp'}}{2}} \, \sqrt{\sin \frac{\pp'}{2}}\Bigg]\ ,
\\\nonumber
& T_1 = - T_2 = - \chi_1^{++} \ ,
\\\nonumber
& Q_1 = Q_2 = \frac{1}{\sin \frac{1}{4}(\pp + \pp')} \Bigg[\chi_2^{++} \, \sin \frac{1}{4}(\pp - \pp') - \, \chi_1^{++} \, \sqrt{\sin \frac{\pp\vphantom{\pp'}}{2}} \,\sqrt{ \sin \frac{\pp'}{2}}\Bigg]\ ,
\\\nonumber
& R_1 = R_2 = \chi_2^{++} \ .
\end{align}
We have checked that the Yangian representation with the coproducts taken in
the appropriate branches -- and away from the bound-state point (see footnote
\ref{footbsp}) -- is fully reducible simultaneously at level zero and one,
which is consistent with the appearance of two undetermined functions in the
scattering matrix. In order to match the limit from the massive S-matrix, the
functions $\chi_{1,2}^{++}$ should be chosen as follows:
\begin{eqnarray}
\chi_2^{++} = -\frac{\sqrt{\sin \frac{\pp\vphantom{\pp'}}{2}} \,\sqrt{ \sin \frac{\pp'}{2}}}{2\sin \frac{1}{4}(\pp + \pp')}\tPo^{++}, \qquad
\chi_1^{++} = \big(\frac{\fl^{++}}2 \, - \, \frac{\sin \frac{1}{4}(\pp - \pp')}{2\sin \frac{1}{4}(\pp + \pp')}\big)\tPo^{++}\ ,
\end{eqnarray}
where $\fl^{++}$ is the limit of $f$. The limit of $f$ is not fixed by the
comparison with the Yangian S-matrix. However, imposing the Yang-Baxter
equation
\begin{eqnarray}\label{ybeppp}
\mathbb{S}^{++}_{12} \, \mathbb{S}^{++}_{13} \, \mathbb{S}^{++}_{23} \, = \, \mathbb{S}^{++}_{23} \, \mathbb{S}^{++}_{13} \, \mathbb{S}^{++}_{12}\ .
\end{eqnarray}
requires that
\begin{eqnarray}
\fl^{++} = \pm 1\,, \, 0\ .
\end{eqnarray}

The Yang-Baxter equation for $\s=\s'=+1$ scattering \eqref{ybeppp} does not
allow for non-constant limits of the function $f$. In particular, the condition
it imposes reads (we denote $\lim_{m,m'\to 0} f(\pp_i,\pp_j) \equiv \fl^{++}_{ij}$)
\begin{eqnarray}
\label{m}
\fl^{++}_{13} - \fl^{++}_{23} + \fl^{++}_{12} \, (\fl^{++}_{13}\fl^{++}_{23}-1) = 0 \ .
\end{eqnarray}
If $\fl^{++}_{13}\fl^{++}_{23}=1$, we immediately get $\fl^{++}=\pm1$. If
$\fl^{++}_{13}\fl^{++}_{23}\neq1$, we find
\begin{eqnarray}
\label{must}
\fl^{++}_{12} = \frac{\fl^{++}_{13} - \fl^{++}_{23}}{1-\fl^{++}_{13}\fl^{++}_{23}}\ .
\end{eqnarray}
However, the l.h.s. of \eqref{must} does not depend on $\pp_3$, and hence we
should impose that the derivative of the r.h.s. with respect to $\pp_3$ is zero.
Doing so, we find that either once again $f^{++}=\pm 1$, or, if $f^{++}\neq\pm
1$, then
\begin{eqnarray}
\label{must2}
\frac{\partial_3 \fl^{++}_{13}}{1 - (\fl^{++}_{13})^2} \, = \,
- \frac{1}{2} \, \partial_3 \, \log\Big(\frac{1 - \fl^{++}_{13}}{1+ \fl^{++}_{13}}\Big)
\end{eqnarray}
should be independent of $\pp_1$. Let us call this function $\omega(\pp_3)$. This
implies that
\begin{eqnarray}
\fl^{++}_{13} = \frac{1 -\bar\omega(\pp_1) \tilde \omega(\pp_3)}{1+\bar \omega (\pp_1) \tilde \omega(\pp_3)}\ ,
\qquad \tilde \omega (\pp_3)=\exp\big[-2\int^{\pp_3} \omega(\pp'_3) d\pp'_3\big]\ .
\end{eqnarray}
Plugging this expression back into \eqref{must} we find that either $\bar
\omega(\pp) = 0$, in which case $f^{++} = 1$ and we are done, or $\bar
\omega(\pp) = \tilde\omega^{-1}(\pp)$. Finally, substituting into \eqref{m}
we find that $\tilde\omega(\pp)$ is a constant and hence $f^{++}=0$. This then
demonstrates that the solutions of \eqref{m} are $f^{++}=\pm 1,0$.

As in the relativistic case \cite{Zamol2}, a different situation applies for
$\s=+1$, $\s'=-1$. The Yangian invariance again fixes the S-matrix
up to two undetermined functions $\chi_{1,2}^{+-}$:
\begin{align}\nonumber
& S_1 = S_2 = \frac{1}{\cos \frac{1}{4}(\pp + \pp')} \Bigg[\chi_1^{+-} \, \cos \frac{1}{4}(\pp - \pp') + \,i \chi_2^{+-} \, \sqrt{\sin \frac{\pp\vphantom{\pp'}}{2}} \, \sqrt{-\sin \frac{\pp'}{2}}\Bigg]\ ,
\\\nonumber
& T_1 = T_2 = \chi_1^{+-} \ ,
\\\nonumber
& Q_1 = Q_2 = \frac{1}{\cos \frac{1}{4}(\pp + \pp')} \Bigg[\chi_2^{+-} \, \cos \frac{1}{4}(\pp - \pp') + \,i \chi_1^{+-} \, \sqrt{\sin \frac{\pp\vphantom{\pp'}}{2}} \,\sqrt{- \sin \frac{\pp'}{2}}\Bigg]\ ,
\\\nonumber
& R_1 = R_2 = \chi_2^{+-} \ .
\end{align}
Again one can check that the Yangian representation with the coproducts taken
in the appropriate branches -- and away from the bound-state point (see
footnote \ref{footbsp}) -- is fully reducible simultaneously at level zero and
one, which is as before consistent with the appearance of two undetermined functions in
the scattering matrix. In order to match the limit from the massive S-matrix,
the functions $\chi_{1,2}^{+-}$ should be chosen as follows:
\begin{eqnarray}
\chi_2^{+-} = -i\frac{\sqrt{\sin \frac{\pp\vphantom{\pp'}}{2}} \, \sqrt{-\sin \frac{\pp'}{2}}}{2\cos \frac{1}{4}(\pp + \pp')}\tPo^{+-}\ , \qquad
\chi_1^{+-} = \big(\frac{\fl^{+-}}{2} \, + \, \frac{\cos \frac{1}{4}(\pp - \pp')}{2\cos \frac{1}{4}(\pp + \pp')}\big)\tPo^{+-}\ ,
\end{eqnarray}
where $\fl^{+-}$ is the limit of $f$. For this mixed case the limit of
$f$ is also not fixed by the comparison with the Yangian S-matrix. Once again, the
Yang-Baxter equation fixes this limiting value. In order to write down the
Yang-Baxter equation for the mixed case, we need to first calculate the S-matrix for
$\s=\s'=-1$, as schematically it is given by
\begin{eqnarray}\label{mixedybe}
\mathbb{S}^{+-}_{12} \, \mathbb{S}^{+-}_{13} \, \mathbb{S}^{--}_{23} \, = \, \mathbb{S}^{--}_{23} \, \mathbb{S}^{+-}_{13} \, \mathbb{S}^{+-}_{12}\ .
\end{eqnarray}

The Yangian invariance again fixes the $\s=\s'=-1$ S-matrix up to two
undetermined functions $\chi_{1,2}^{--}$:
\begin{align}\nonumber
& S_1 = -S_2 = \frac{1}{\sin \frac{1}{4}(\pp + \pp')} \Bigg[\chi_1^{--} \, \sin \frac{1}{4}(\pp - \pp') - \, \chi_2^{--} \, \sqrt{-\sin \frac{\pp\vphantom{pp'}}{2}} \, \sqrt{-\sin \frac{\pp'}{2}}\Bigg]\ ,
\\\nonumber
& T_1 = - T_2 = - \chi_1^{--} \ ,
\\\nonumber
& Q_1 = Q_2 = \frac{1}{\sin \frac{1}{4}(\pp + \pp')} \Bigg[-\chi_2^{--} \, \sin \frac{1}{4}(\pp - \pp') - \, \chi_1^{--} \, \sqrt{-\sin \frac{\pp\vphantom{pp'}}{2}} \,\sqrt{- \sin \frac{\pp'}{2}}\Bigg]\ ,
\\\nonumber
& R_1 = R_2 = \chi_2^{--} \ .
\end{align}
In order to match the limit from the massive S-matrix, the functions
$\chi_{1,2}^{--}$ have to be chosen as follows:
\begin{eqnarray}
\chi^{--}_2 = \frac{\sqrt{-\sin \frac{\pp\vphantom{\pp'}}{2}} \, \sqrt{-\sin \frac{\pp'}{2}}}{2\sin \frac{1}{4}(\pp + \pp')}\tPo^{--}\ , \qquad
\chi^{--}_1 = \big(-\frac{\fl^{--}}2 \, - \, \frac{\sin \frac{1}{4}(\pp - \pp')}{2\sin \frac{1}{4}(\pp + \pp')}\big)\tPo^{--}\ ,
\end{eqnarray}
where $\fl^{--}$ is the limit of $f$. The Yang-Baxter equation
\begin{eqnarray}
\mathbb{S}^{--}_{12} \, \mathbb{S}^{--}_{13} \, \mathbb{S}^{--}_{23} \, = \, \mathbb{S}^{--}_{23} \, \mathbb{S}^{--}_{13} \, \mathbb{S}^{--}_{12}\ .
\end{eqnarray}
fixes this limiting value to
\begin{eqnarray}
\fl^{--} = \pm 1\,, \, 0\ .
\end{eqnarray}
Taking this result into account, the mixed Yang-Baxter equation
\eqref{mixedybe} fixes $\fl^{+-}=\pm 1$ if one chooses either $\fl^{--}=1$ or
$\fl^{--}=-1$, or $\fl^{+-}$ to any constant if one chooses $\fl^{--}=0$.

To exhaust all possibilities, the $\s=-1,\,\s'=+1$ S-matrix is given by
\begin{align}\nonumber
& S_1 = S_2 = \frac{1}{\cos \frac{1}{4}(\pp + \pp')} \Bigg[\chi_1^{-+} \, \cos \frac{1}{4}(\pp - \pp') - \,i \chi_2^{-+} \, \sqrt{-\sin \frac{\pp\vphantom{\pp'}}{2}} \,\sqrt{ \sin \frac{\pp'}{2}}\Bigg]\ ,
\\\nonumber
& T_1 = T_2 = \chi_1^{-+} \ ,
\\\nonumber
& Q_1 = Q_2 = \frac{1}{\cos \frac{1}{4}(\pp + \pp')} \Bigg[-\chi_2^{-+} \, \cos \frac{1}{4}(\pp - \pp') + \,i \chi_1^{-+} \, \sqrt{-\sin \frac{\pp\vphantom{\pp'}}{2}} \,\sqrt{ \sin \frac{\pp'}{2}}\Bigg]\ ,
\\\nonumber
& R_1 = R_2 = \chi_2^{-+} \ .
\end{align}
In order to match the limit from the massive S-matrix, the functions
$\chi_{1,2}^{-+}$ have to be chosen as follows:
\begin{eqnarray}
\chi_2^{-+} = i\frac{\sqrt{-\sin \frac{\pp\vphantom{\pp'}}{2}} \,\sqrt{ \sin \frac{\pp'}{2}}}{2\cos \frac{1}{4}(\pp + \pp')}\tPo^{-+}\ , \qquad
\chi_1^{-+} = \big(- \frac{\fl^{-+}}2 \, + \, \frac{\cos \frac{1}{4}(\pp - \pp')}{2\cos \frac{1}{4}(\pp + \pp')}\big)\tPo^{-+}\ ,
\end{eqnarray}
where $\fl^{-+}$ is the limit of $f$.

By imposing the Yang-Baxter equation for all possible remaining sequences of
scattering processes we find the following possibilities for
the limits of $f$:
\begin{alignat}{4}
&\fl^{++}=\pm1,\,0 \ ,&\qquad&\fl^{+-}= \pm 1\ , &\qquad& \fl^{-+}=\pm 1\ , &\qquad& \fl^{--} = \pm1,\,0 \ ,\no
\\& \fl^{++} = 0\ , &\qquad&\fl^{+-}= \mu_1 , &\qquad& \fl^{-+}=\mu_2\ , &\qquad& \fl^{--} = 0 \ ,
\end{alignat}
where $\mu_1$ and $\mu_2$ are arbitrary constants.
\corr{Note that we have not included the following 
two Yang-Baxter equations:
\begin{equation}\begin{split}
\mathbb{S}_{12}^{+-}\,\mathbb{S}_{13}^{++}\,\mathbb{S}_{23}^{-+}\, =\, \mathbb{S}_{23}^{-+}\,\mathbb{S}_{13}^{++}\,\mathbb{S}_{12}^{+-}\ , \qquad
\mathbb{S}_{12}^{-+}\,\mathbb{S}_{13}^{--}\,\mathbb{S}_{23}^{+-}\, =\, \mathbb{S}_{23}^{+-}\,\mathbb{S}_{13}^{--}\,\mathbb{S}_{12}^{-+}\ ,
\end{split}\end{equation}
as they do not correspond to physically realizable scattering processes. If
particles $1$ and $3$ are both right- or left-moving then they have to scatter
with each other before scattering with an excitation travelling in the opposite
direction. If we formally include them then the possibilities for the limits of
$f$ are reduced to
\begin{eqnarray}
(\fl^{++},\fl^{+-},\fl^{-+},\fl^{--}) \,&\in & \,
\{ (1,1,1,1) \, , \, (-1,-1,-1,-1) \, , \, (0,\mu,-\mu,\tilde\mu)\, , \, (\tilde\mu,\mu,-\mu,0)\}\ ,
\end{eqnarray}
with $\mu$ any constant for $\tilde \mu =0$, $\m=\pm 1$ for $\tilde \mu=1$, and
$\mu=\pm1$ for $\tilde \mu=-1$.}

The various choices for $\fl^{++},\fl^{+-},\fl^{-+},\fl^{--}$ can be further
restricted by considering crossing symmetry. Although in the massless case
there is no clear physical interpretation of crossing, see, for example,
\cite{Zamol2}, one may nevertheless demand that it is still present. Let us
recall that the crossing transformation simultaneously changes the sign of the
energy and momentum, therefore the crossing of a $+$ ($-$) particle is
still a $+$ ($-$) particle. Consequently in the crossing relation
\eqref{matricial} we should consider two massless S-matrices of the same type.
Considering the various possible limits of $f$, we find that the choices
$\fl^{++}=0$ and $\fl^{--}=0$ are incompatible with crossing. Indeed, before
taking the massless limit, the function $f$ satisfies the following crossing
transformation with respect to the first particle:
\begin{eqnarray}
f \to \frac{x'^+ \, x'^-}{f}\ ,
\end{eqnarray}
which is is clearly problematic for $f\to0$. We are then left with the
following choices for the limits of $f$
\begin{eqnarray}
\label{2ch}
\fl^{++}=\pm1 \ , \qquad \fl^{+-}= \pm 1\ , \qquad \fl^{-+}=\pm 1\ , \qquad \fl^{--} = \pm1 \ .
\end{eqnarray}
It is worth noting that for the crossing relation to be satisfied for these
choices we should not only consider two massless S-matrices of the same type,
but also with the same limit of $f$.

Now that we are left with the choices in eq.~\eqref{2ch}, let us recall that in
the massive case the sign of $f$ is not determined by symmetry or the
Yang-Baxter equation, rather from comparing with perturbation theory. This is
consistent with the residual ambiguity we are finding in this limit.

If we look at the BMN limit (see section \ref{secpert}) for the $\s= \s' =\pm1$
S-matrices, we don't necessarily expect to (and indeed we do not) find the
identity. This expectation comes from the fact that the quadratic Lagrangian of
the light-cone gauge-fixed theory is relativistic and it is not clear how one
should perform a perturbative computation for the scattering of two massless
relativistic particles on the same branch, or if there should be a perturbative
expansion at all.

For the $\s=-\s'=\pm1$ S-matrices one may expect the limit to be better behaved
as perturbative computations can be carried out. Indeed, assuming that the
phase goes like one plus corrections, then for the $\s=-\s'=+1$ case we
find that if $f^{+-} = 1$ the S-matrix is the identity at leading order, while
for the $\s=-\s'=-1$ case the same is true, but with $f^{-+} = -1$. Therefore,
we end up with the following choices for the limits of $f$
\begin{eqnarray}
\label{fc}
\fl^{++}=\pm1 \ , \qquad \fl^{+-}= 1\ , \qquad \fl^{-+}=- 1\ , \qquad \fl^{--} = \pm1 \ .
\end{eqnarray}

We may attribute some physical meaning to this result by considering the group
velocities
\begin{equation}
\vv = \frac{\partial{\ee}}{\partial{\pp}} \ ,\qquad
\vv' = \frac{\partial{\ee'}}{\partial{\pp'}} \ .
\end{equation}
{\color{black} Let us remark that our considerations (especially those referring to the ordering of velocities) will only apply when trying to attach a physical interpretation of real time scattering to these amplitudes.
In general, for a complete analysis, one should also consider the possibility of analytically continuing the S-matrices as functions of the kinematical variables.
With this in mind, for} a physically realizable scattering process with $\s =-\s'=+1$ the group
velocities satisfy $\vv>\vv'$, while for a scattering process with $\s =
-\s'=-1$ we have $\vv'>\vv$.  Therefore, we may associate $\lim_{m,m'\to 0} f
\to 1$ with $\vv>\vv'$ and $\lim_{m,m'\to 0}f\to -1$ with $\vv<\vv'$. This is
consistent with the crossing symmetry discussed above as the group velocity is
invariant under the crossing transformation. Furthermore, one may expect the
$\s=-\s'=+1$ and $\s=-\s'=-1$ S-matrices to be related upon interchanging the
arguments. Indeed, 
the following equation is
satisfied for real momenta\,\footnote{Here we are defining $\mathbb{S}
|\Phi^{\vphantom{'}}_a\Phi'_b\ket =
\mathcal{S}_{ab}^{cd}(\pp,\pp')|\Phi^{\vphantom{'}}_c\Phi'_d\ket$, $\Phi_0
=\phi$, $\Phi_1=\psi$ and $[a]=a$.\fnsv}
\begin{equation}\label{ha}
\mathcal{S}^{\pm\mp}{}_{ab}^{cd}(\pp,\pp')\big|_{f\to \pm1} = (-1)^{[a][b]+[c][d]} \mathcal{S}^{\mp\pm}{}_{ba}^{dc}(\pp',\pp)^*\big|_{f\to \mp 1}\ .
\end{equation}
The corresponding relation for the $\s=\s'=\pm1$ S-matrices is given by
\begin{equation}\label{ha2}
\mathcal{S}^{\pm\pm}{}_{ab}^{cd}(\pp,\pp')\big|_{f\to \pm1} = (-1)^{[a][b]+[c][d]} \mathcal{S}^{\pm\pm}{}_{ba}^{dc}(\pp',\pp)^*\big|_{f\to \mp 1}\ .
\end{equation}

To conclude, let us briefly comment on unitarity. Motivated by the
physical interpretation outlined above, one may expect that braiding unitarity
for the massless S-matrix will involve one S-matrix with $f \to 1$ and one with
$f \to -1$, and indeed, one can explicitly check that braiding unitarity
relations can be constructed in this way. They are given by
\begin{equation}\begin{split}\label{masles}
& (-1)^{[c][d]+[e][f]}\mathcal{S}^{\pm\pm}{}_{ab}^{ef}(\pp,\pp')\big|_{f\to \pm1} \mathcal{S}^{\pm\pm}{}_{fe}^{dc}(\pp',\pp)\big|_{f\to \mp 1}\propto\d_a^c\d_b^d\ .
\\
& (-1)^{[c][d]+[e][f]}\mathcal{S}^{\pm\mp}{}_{ab}^{ef}(\pp,\pp')\big|_{f\to \pm1} \mathcal{S}^{\mp\pm}{}_{fe}^{dc}(\pp',\pp)\big|_{f\to \mp 1}\propto\d_a^c\d_b^d\ .
\end{split}\end{equation}
These relations can also be found by taking the massless limit of the braiding
unitarity relation for the massive S-matrix. Finally, one can see that by
combining \eqref{ha}, \eqref{ha2} and \eqref{masles}, all the four massless
S-matrices are also QFT unitary so long as the overall factors satisfy
appropriate constraints.

\subsection{Massless limits and symmetry enhancement}

Let us now consider taking the various massless limits of the parametrizing
functions of the massive S-matrix, {\em i.e.} one massless and one massive or
two massive particles. Here we work in terms of the variables $x^\pm$,
$x'^\pm$ as it allows us to consider the four cases of section \ref{maya} at
the same time. For convenience we introduce the following notation for the
massless Zhukovsky variables
\begin{equation}
\xx = x^+ = \frac{1}{x^-}\ ,\qquad
\xx' = x'^+ = \frac{1}{x'^-}\ .
\end{equation}
The parametrizing functions are then given by
{\allowdisplaybreaks\begin{align}
&\textbf{Massive-Massless} \qquad \qquad \qquad f \to x^-\sqrt{\frac{x^+}{x^-}} \no
\\
& S_1 = T_1 = - \combop\frac{(x^+-\xx')+\sqrt{\frac{x^+}{x^-}}(x^--\xx')}{2(1-x^+\xx')}\tPo \ ,
& & S_2 = T_2 = \frac{(1-x^+\xx')+\sqrt{\frac{x^+}{x^-}}(1-x^-\xx')}{2(1-x^+\xx')}\tPo \ ,\no
\\
& Q_1 = Q_2 = i \sqrt[4]{\frac{x^+}{x^-}\frac1{\xx'^{2}}}\combop \frac{\xx'\eta\eta'}{2(1-x^+\xx')}\tPo\ ,
& & R_1 = R_2 = i \sqrt[4]{\frac{x^+}{x^-}\frac1{\xx'^{2}}} \frac{\xx'\eta\eta'}{2(1-x^+\xx')}\tPo\ ,\label{d1}
\\ &&&\no \\
&\textbf{Massless-Massive} \qquad \qquad \qquad f \to -x'^-\sqrt{\frac{x'^+}{x'^-}}\no
\\
& S_1 = T_2 = \combo\frac{(1-\xx x'^-)+\sqrt{\frac{x'^-}{x'^+}}(1-\xx x'^+)}{2(\xx-x'^-)}\tPo \ ,
&& S_2 = T_1 =\frac{(\xx-x'^-)+ \sqrt{\frac{x'^-}{x'^+}}(\xx-x'^+)}{2(\xx-x'^-)}\tPo \ ,\no
\\
& Q_1 = Q_2 = i \sqrt[4]{\xx^2\frac{x'^-}{x'^+}}\combo\frac{\eta\eta'}{2(\xx-x'^-)}\tPo\ ,
&& R_1 = R_2 = - i \sqrt[4]{\xx^2\frac{x'^-}{x'^+}}\frac{\eta\eta'}{2(\xx-x'^-)}\tPo\ ,\label{d2}
\\ &&&\no \\
&\textbf{Massless-Massless} \qquad \qquad \qquad f \to \pm 1\no
\\
& S_1 = - \combo\combop\frac{1-\xx \xx' \pm(\xx - \xx')}{2(1-\xx\xx')}\tPo\ , \qquad
&& S_2 =\frac{1-\xx \xx' \pm(\xx - \xx')}{2(1-\xx\xx')}\tPo\ ,\no
\\
& T_1 = - \combop\frac{\xx - \xx'\pm(1-\xx\xx')}{2(1-\xx\xx')}\tPo\ , \qquad
&& T_2 = \combo\frac{\xx - \xx'\pm(1-\xx\xx')}{2(1-\xx\xx')}\tPo\ ,\no
\\
& Q_1 = Q_2 =\pm i\combo\combop\sqrt[4]{\frac{\xx^2}{\xx'^2}} \frac{\xx'\eta\eta'}{2(1-\xx\xx')}\tPo\ ,
& & R_1 = R_2 = i\sqrt[4]{\frac{\xx^2}{\xx'^2}} \frac{\xx'\eta\eta'}{2(1-\xx\xx')}\tPo\ .\label{d3}
\end{align}}%
Given that $\combo$ and $\combop$ are equal to $\pm 1$ one can see that the
limit of the function $f$ is well-defined if we just take one of the two masses
to zero. In particular, taking $m \to0$ we have $f \to x^-\sqrt{\frac{x^+}{x^-}}$
while for $m'\to0$ we have $f \to -x'^-\sqrt{\frac{x'^+}{x'^-}}$.

The factors of $\combo$ and $\combop$ in \eqref{d3} are the origin of the
various expressions for the different choices of $\sigma$ and $\sigma'$ in
section \ref{maya}. For example, to recover the results of section \ref{maya}
we should take $\combo = 1$ for $\s=+1$ and $\combo = -1$ for $\s=-1$, and
similarly for $\xx'$. For $\pp \in [-\pi,\pi]$, this again corresponds to
taking the branch cut on the negative real axis.

We may also consider taking the massless limit of the S-matrices for one
massive and one massless excitation. Following the same set of rules as above,
{\em i.e.} setting $\combo$ equal to $1$ for $\s=+1$ and $-1$ for $\s=-1$, and
similarly for $\combop$, the following table gives the expressions we find for
the limits of $f$
\begin{center}
\begin{tabular}{ccl}
\hline\hline
Before limit & After limit & Limit of $f$
\\\hline\hline
Massive - Massless ($\s'=+1$) \qquad & Massless-Massless ($\s=+1$, $\s'=+1$) \qquad & $f^{++} = 1$
\\ Massive - Massless ($\s'=+1$) \qquad & Massless-Massless ($\s=-1$, $\s'=+1$) \qquad & $f^{-+}= -1$
\\ Massive - Massless ($\s'=-1$) \qquad & Massless-Massless ($\s=+1$, $\s'=-1$) \qquad & $f^{+-}= 1$
\\ Massive - Massless ($\s'=-1$) \qquad & Massless-Massless ($\s=-1$, $\s'=-1$) \qquad & $f^{--}= -1$
\\ Massless - Massive ($\s=+1$) \qquad & Massless-Massless ($\s=+1$, $\s'=+1$) \qquad & $f^{++} = - 1$
\\ Massless - Massive ($\s=+1$) \qquad & Massless-Massless ($\s=+1$, $\s'=-1$) \qquad & $f^{+-} =1$
\\ Massless - Massive ($\s=-1$) \qquad & Massless-Massless ($\s=-1$, $\s'=+1$) \qquad & $f^{-+} =- 1$
\\ Massless - Massive ($\s=-1$) \qquad & Massless-Massless ($\s=-1$, $\s'=-1$) \qquad & $f^{--} = 1$
\\\hline\hline
\end{tabular}
\end{center}
Therefore we find the same set of possible limits of $f$ as found from the
analysis in section \ref{maya}, the result of which is given in eq.~\eqref{fc}.

Finally, from eqs.~\eqref{d1}--\eqref{d3} we can see that taking the various
massless limits results in many of the parametrizing functions (or products
thereof) coinciding. It is clear from the expressions in appendix \ref{appu1}
that there will then be additional $U(1)$ symmetries of the S-matrix acting on
both the bosons and fermions. This is surely required for these S-matrices to
describe the scattering of the massless modes of the light-cone gauge $AdS_2
\times S^2 \times T^6$ superstring as they (the bosons and fermions) will
transform under various $U(1)$ symmetries originating from the $T^6$
compact space \cite{amsw}. The precise construction of the S-matrices involving
massless modes from the building blocks described above requires the knowledge
of the full light-cone gauge symmetry algebra and its action on all the states,
as was done for $AdS_3 \times S^3 \times T^4$ in \cite{BogdanLatest} and $AdS_5
\times S^5$ in \cite{Arutyunov:2006ak}.

\section{Bethe Ansatz}\label{secu1}

As discussed at the beginning of section \ref{sec1} the tensor product of two
copies of any S-matrix of the form \eqref{ans} satisfying \eqref{trdeteq}
possesses an additional $U(1)$ symmetry, which does not have a well-defined
action on the individual factor S-matrices. This symmetry is expected from
string theory as a consequence of the additional compact space $T^6$ required
for a consistent 10-d superstring theory \cite{amsw}.\footnote{We are grateful
to O.~Ohlsson Sax and P.~Sundin for pointing out to us the existence of this
symmetry in the superstring theory.\fnsv} Under this symmetry the bosons $y$
and $z$ are uncharged, while the fermions $(\z,\chi)^T$ form an $SO(2)$ vector.
Furthermore $(\mathfrak{Q}_2,\mathfrak{Q}_1)^T$ and
$(\mathfrak{S}_2,\mathfrak{S}_1)^T$ are also charged as $SO(2)$ vectors under
the symmetry.\footnote{Here the subscripts on the supercharges $\mathfrak{Q}$
and $\mathfrak{S}$ refer to the two copies of $\mathfrak{psu}(1|1)$ in the full
symmetry algebra. In particular the charges with the label $1$ act on the first
entry in the tensor product \eqref{bhtens}, while the charges with the label
$2$ on the second entry.\fnsv}

Here we will summarize the relevant details of this symmetry. Explicit details
(including the expansion of the tensor product) are given in appendix
\ref{appu1}. Defining
\begin{equation}
|\qpm\ket = \frac{1}{\sqrt{2}} (|\z\ket \pm i |\chi\ket) \ , \qquad
\mathfrak{G}_\mathfrak{q\pm} = \frac{1}{\sqrt{2}} ( \mathfrak{Q}_2 \pm i \mathfrak{Q}_1) \ , \qquad
\mathfrak{G}_\mathfrak{s\pm} = \frac{1}{\sqrt{2}} ( \mathfrak{S}_2 \pm i \mathfrak{S}_1) \ ,
\end{equation}
and their conjugates, we have the following actions of the $U(1)$ generator,
$\mathfrak{J}_{_{U(1)}}$,
\begin{equation}
\mathfrak{J}_{_{U(1)}}|\qpm\ket = \pm i |\qpm\ket \ , \qquad
[\mathfrak{J}_{_{U(1)}}, \mathfrak{G}_{\mathfrak{q,s}\pm}] = \pm i \mathfrak{G}_{\mathfrak{q,s}\pm} \ .
\end{equation}

\

To proceed with the algebraic Bethe ansatz (ABA) technique one constructs
the monodromy matrix as a string of R-matrices acting on an auxiliary space $a$
and on $N$ physical spaces
\begin{eqnarray}
\label{ABCD}
T_a (\lambda) = R_{a,1} \cdot \, ... \, \cdot R_{a,N} \, = \, \begin{pmatrix}A(\lambda)&B(\lambda)\\C(\lambda)&D(\lambda)
\end{pmatrix}\ ,
\end{eqnarray}
where $\cdot$ denotes multiplication in the auxiliary space. $A(\lambda)$,
$B(\lambda)$, $C(\lambda)$ and $D(\lambda)$ are operators on $N$-particle
physical space, while the $2\times 2$ matrix acts on the auxiliary space. As a
consequence of the Yang-Baxter equation one has
\begin{eqnarray}
\label{RTT}
R_{a_1,a_2} (\lambda_1 - \lambda_2) \, T_{a_1} (\lambda_1)\, T_{a_2} (\lambda_1)\,
= \, T_{a_2} (\lambda_1)\, T_{a_1} (\lambda_1)\, R_{a_1,a_2} (\lambda_1 - \lambda_2)\ .
\end{eqnarray}
Taking the trace $tr_{a_1} \otimes tr_{a_2}$ on both sides of \eqref{RTT}, one
finds that the {\it transfer matrix} $T(\lambda) \equiv tr \, T_a (\lambda) =
A(\lambda) + D(\lambda)$ satisfies:
\begin{eqnarray}
\label{being}
[T(\lambda), T(\lambda')]=0\ .
\end{eqnarray}
As $T(\lambda)$ is an $N$th order polynomial in $\lambda$ (with the highest-power
coefficient chosen equal to 1), we see that \eqref{being} implies that $T(\lambda)$
generates $N$ non-trivial independent commuting operators.

To find the simultaneous eigenvectors of all the commuting charges (which
include the Hamiltonian), one assumes that $B (\lambda)$ is a creation operator
acting on a pseudo-vacuum $|vac\rangle$, which is annihilated by $C(\lambda)$:
\begin{eqnarray}
\label{BBBB}
|\Psi (\lambda_1,...,\lambda_M) \rangle = B (\lambda_1) ... B (\lambda_M) \, |vac\rangle\ .
\end{eqnarray}
The pseudo-vacuum should be a highest-weight $T(\lambda)$-eigenstate, whether
or not that is the true ground state of the Hamiltonian. The vectors
\eqref{BBBB} are not immediately eigenstates of $T(\lambda)$ because of
unwanted terms obtained when acting with $T(\lambda)$. These unwanted terms are
cancelled by imposing the Bethe equations, providing the quantization condition
for the momenta of excitations.

Let us now give some initial observations on applying the ABA procedure to the
S-matrix for the light-cone gauge $AdS_2 \times S^2 \times T^6$ superstring.
We can immediately remark that a single copy of the centrally-extended S-matrix
does not seem to admit a pseudovacuum on which to construct the ABA procedure.
However, when we take the tensor product of two copies there is a pseudovacuum.
This is given by a uniform sequence of either all $|\theta_+\rangle$ states or,
alternatively, $|\theta_-\rangle$. In fact, thanks to the conservation of the
additional $U(1)$ charge discussed above and in appendix \ref{appu1}, these
states are the only ones with maximal (minimal) such charge, and therefore have
to be eigenvalues of the transfer matrix. By a similar logic they are also
annihilated by some of the lower-corner entries of the (now 4-dimensional)
transfer matrix. This in principle could allow the ABA procedure to be applied.
However, this still remains technically challenging given the complexity of the
parametrizing functions of the S-matrix.

\section{Comments}\label{comments}

In this paper we have constructed the S-matrix describing the scattering of
{\color{black} particular representations of the centrally-extended \corp{$\alg{psu}(1|1)^2$}
Lie superalgebra, conjectured to be related to the} massive modes of the $AdS_2 \times S^2 \times T^6$ light-cone gauge
superstring. A significant difference with the $AdS_5 \times S^5$ and $AdS_3
\times S^3\times T^4$ light-cone gauge superstrings is that the massive
excitations {\color{black} are taken to} transform in long representations of the symmetry algebra
$\mathfrak{psu}(1|1)^2\ltimes \mathbb{R}^3$. Consequently, {\color{black} under these assumptions} there is no
shortening condition and the dispersion relation is not {\color{black} entirely} fixed by symmetry.
Furthermore, the symmetry only fixes the S-matrix up to an overall phase, for
which we have given the crossing and unitarity relations, which
appear to be more complicated than those in the $AdS_5 \times S^5$ case.
The exact form of both the dispersion relation and the phase remain
to be determined.

{\color{black} We have identified a natural way to take the massless limit on these representations, and have analyzed in detail the limits} (one massive and one massless or two massless particles) of
the massive S-matrix. The resulting expressions
should play the role of building blocks for the S-matrices of the massless
modes of the $AdS_2 \times S^2 \times T^6$ superstring. As for the $AdS_3
\times S^3 \times T^4$ case \cite{BogdanLatest}, the precise nature of this
construction requires the knowledge of how all the states transform under the
full light-cone gauge symmetry algebra including any additional bosonic
symmetries originating from the $T^6$ compact directions.

In the massless limit the light-cone gauge symmetry $\mathfrak{psu}(1|1)^2
\ltimes \mathbb{R}^3$ can be extended to a Yangian of the standard form.
However this does not generalize in an obvious way to the massive S-matrix. It
would be interesting to see if there exists a non-standard Yangian in this
case. We are also currently investigating the presence of the secret symmetry
\cite{Matsumoto:2007rh,Beisert:2007ty} and the RTT realization of the symmetry
algebra \cite{Beisert:2014hya,Pittelli:2014ria}. Finally, we gave some initial
considerations regarding the Bethe ansatz for the massive S-matrix, in
particular highlighting the existence of a pseudovacuum. Due to the complexity
of the parametrizing functions of the S-matrix and the fact that we are
considering long representations of the symmetry algebra the completion of the
algebraic Bethe ansatz remains an open problem.

\section*{Acknowledgments}

We would like to thank R.~Borsato, A.~Sfondrini, B.~Stefanski, O.~Ohlsson Sax,
R.~Roiban, P.~Sundin, A.~Tseytlin, M.~Wolf and L.~Wulff for related discussions
and A.~Tseytlin for useful comments on the draft. B.H. would like to thank
A.~Tseytlin in particular for interesting discussions in the early stages of
this project. B.H. is supported by the Emmy Noether Programme ``Gauge fields
from Strings'' funded by the German Research Foundation (DFG). A.P. is
supported in part by the EPSRC under the grant EP/K503186/1. A.T. thanks EPSRC
for funding under the First Grant project EP/K014412/1 ``Exotic quantum groups,
Lie superalgebras and integrable systems".

\appendix

\refstepcounter{section}
\def\theequation{A.\arabic{equation}}
\setcounter{equation}{0}
\section*{Appendix A: Expansion of tensor product and $\mathbf{U(1)}$ symmetry}\label{appu1}
\addcontentsline{toc}{section}{A \ Expansion of tensor product and \texorpdfstring{$\mathbf{U(1)}$}{U(1)} symmetry}

In this appendix we will write explicitly the full expression for the tensor
product of two copies of the S-matrix given in \eqref{ans}. This will allow us
to demonstrate the existence of the $U(1)$ symmetry that was important in
section \ref{secu1} for the Bethe ansatz.
\allowdisplaybreaks{\begin{align}
& \textbf{Boson-Boson} \no
\\ & \mathbb{S} |y y'\ket = S_1^2 |y y'\ket - Q_1^2 |z z'\ket + S_1Q_1 ( |\z\z'\ket + |\c\c'\ket)\no
\\ & \mathbb{S} |z z'\ket = S_2^2 |z z'\ket - Q_2^2 |y y'\ket - S_2Q_2 ( |\z\z'\ket + |\c\c'\ket)\no
\\ & \mathbb{S} |y z'\ket = T_1^2 |y z'\ket + R_1^2 |z y'\ket - T_1R_1 ( |\z\c'\ket - |\c\z'\ket)\no
\\ & \mathbb{S} |z y'\ket = T_2^2 |z y'\ket + R_2^2 |y z'\ket - T_2R_2 ( |\z\c'\ket - |\c\z'\ket)\no
\\& \textbf{Boson-Fermion} \no
\\ & \mathbb{S} |y \z'\ket = S_1T_1 |y \z'\ket - Q_1R_1 |z \c'\ket + S_1R_1 |\z y'\ket + T_1Q_1|\c z'\ket\no
\\ & \mathbb{S} |y \c'\ket = S_1T_1 |y \c'\ket + Q_1R_1 |z \z'\ket + S_1R_1 |\c y'\ket - T_1Q_1|\z z'\ket\no
\\ & \mathbb{S} |z \z'\ket = S_2T_2 |z \z'\ket + Q_2R_2 |y \c'\ket - S_2R_2 |\z z'\ket + T_2Q_2|\c y'\ket\no
\\ & \mathbb{S} |z \c'\ket = S_2T_2 |z \c'\ket - Q_2R_2 |y \z'\ket - S_2R_2 |\c z'\ket - T_2Q_2|\z y'\ket\no
\\& \textbf{Fermion-Boson} \no
\\ & \mathbb{S} |\z y'\ket = S_1T_2 |\z y'\ket + Q_1R_2 |\c z'\ket + S_1R_2 |y\z'\ket - T_1Q_2|z\c'\ket\no
\\ & \mathbb{S} |\c y'\ket = S_1T_2 |\c y'\ket - Q_1R_2 |\z z'\ket + S_1R_2 |y\c'\ket + T_1Q_2|z\z'\ket\no
\\ & \mathbb{S} |\z z'\ket = S_2T_1 |\z z'\ket - Q_2R_1 |\c y'\ket - S_2R_1 |z\z'\ket - T_2Q_1|y\c'\ket\no
\\ & \mathbb{S} |\c z'\ket = S_2T_1 |\c z'\ket + Q_2R_1 |\z y'\ket - S_2R_1 |z\c'\ket + T_2Q_1|y\z'\ket\no
\\& \textbf{Fermion-Fermion} \no
\\ & \mathbb{S} |\z \z'\ket = S_1S_2 |\z \z'\ket + Q_1Q_2 |\c \c'\ket + S_1Q_2 |yy'\ket - S_2Q_1|zz'\ket\no
\\ & \mathbb{S} |\c \c'\ket = S_1S_2 |\c \c'\ket + Q_1Q_2 |\z \z'\ket + S_1Q_2 |yy'\ket - S_2Q_1|zz'\ket\no
\\ & \mathbb{S} |\z \c'\ket = T_1T_2 |\z \c'\ket - R_1R_2 |\c \z'\ket - T_1R_2 |yz'\ket - T_2R_1|zy'\ket\no
\\ & \mathbb{S} |\c \z'\ket = T_1T_2 |\c \z'\ket - R_1R_2 |\z \c'\ket + T_1R_2 |yz'\ket + T_2R_1|zy'\ket
\end{align}}

Let us now perform a change of basis for the fermionic states
\begin{equation}
|\theta_\pm\ket = \frac{1}{\sqrt{2}}(|\z\ket \pm i |\c\ket)\ ,
\end{equation}
such that in this basis the S-matrix has the form
\allowdisplaybreaks{\begin{align}
& \textbf{Boson-Boson} \no
\\ & \mathbb{S} |y y'\ket = S_1^2 |y y'\ket - Q_1^2 |z z'\ket + S_1Q_1 ( |\qp\qm'\ket + |\qm\qp'\ket)\no
\\ & \mathbb{S} |z z'\ket = S_2^2 |z z'\ket - Q_2^2 |y y'\ket - S_2Q_2 ( |\qp\qm'\ket + |\qm\qp'\ket)\no
\\ & \mathbb{S} |y z'\ket = T_1^2 |y z'\ket + R_1^2 |z y'\ket - i T_1R_1 ( |\qp\qm'\ket - |\qm\qp'\ket)\no
\\ & \mathbb{S} |z y'\ket = T_2^2 |z y'\ket + R_2^2 |y z'\ket - i T_2R_2 ( |\qp\qm'\ket - |\qm\qp'\ket)\no
\\& \textbf{Boson-Fermion} \no
\\ & \mathbb{S} |y \qpm'\ket = S_1T_1 |y \qpm'\ket \pm i Q_1R_1 |z \qpm'\ket + S_1R_1 |\qpm y'\ket \mp i T_1Q_1|\qpm z'\ket\no
\\ & \mathbb{S} |z \qpm'\ket = S_2T_2 |z \qpm'\ket \mp i Q_2R_2 |y \qpm'\ket - S_2R_2 |\qpm z'\ket \mp i T_2Q_2|\qpm y'\ket\no
\\& \textbf{Fermion-Boson} \no
\\ & \mathbb{S} |\qpm y'\ket = S_1T_2 |\qpm y'\ket \mp i Q_1R_2 |\qpm z'\ket + S_1R_2 |y\qpm'\ket \pm i T_1Q_2|z\qpm'\ket\no
\\ & \mathbb{S} |\qpm z'\ket = S_2T_1 |\qpm z'\ket \pm i Q_2R_1 |\qpm y'\ket - S_2R_1 |z\qpm'\ket \pm i T_2Q_1|y\qpm'\ket\no
\\& \textbf{Fermion-Fermion} \no
\\ & \mathbb{S} |\qpm \qmp'\ket = \tfrac12 (S_1S_2 + Q_1Q_2 + T_1T_2 + R_1R_2)|\qpm\qmp'\ket \no
+ \tfrac12 (S_1S_2 + Q_1Q_2 - T_1T_2 - R_1R_2)|\qmp\qpm'\ket\no
\\ & \qquad \qquad \quad + S_1Q_2 |yy'\ket - S_2 Q_1 |zz'\ket \pm i T_1R_2 |yz'\ket \pm i T_2R_1 |zy'\ket\no
\\ & \mathbb{S} |\qpm \qpm'\ket = \tfrac12 (S_1 S_2 - Q_1 Q_2 + T_1 T_2 - R_1 R_2) |\qpm\qpm'\ket
+ \tfrac12 (S_1 S_2 - Q_1 Q_2 - T_1 T_2 + R_1 R_2) |\qmp\qmp'\ket
\end{align}}
Provided that
\begin{equation}
S_1 S_2 - Q_1 Q_2 = T_1 T_2 - R_1 R_2 \ ,
\end{equation}
which was indeed the case for the S-matrix under consideration in the main text
\eqref{trdeteq}, it is clear that this S-matrix commutes with a $U(1)$ symmetry
acting on the states as follows
\begin{equation}
\mathfrak{J}_{_{U(1)}}|y\ket = 0 \ , \qquad \mathfrak{J}_{_{U(1)}}|z\ket = 0\ , \qquad
\mathfrak{J}_{_{U(1)}}|\qpm\ket = \pm i |\qpm \ket \ .
\end{equation}

Finally for completeness we give the commutation relations of the full algebra
under which the S-matrix is invariant. First let us define
\begin{equation}
\mathfrak{G}_\mathfrak{q\pm} = \frac{1}{\sqrt{2}} ( \mathfrak{Q}_2 \pm i \mathfrak{Q}_1) \ , \qquad
\mathfrak{G}_\mathfrak{s\pm} = \frac{1}{\sqrt{2}} ( \mathfrak{S}_2 \pm i \mathfrak{S}_1) \ ,
\end{equation}
where the subscripts on the supercharges $\mathfrak{Q}$ and $\mathfrak{S}$
refer to the two copies of $\mathfrak{psu}(1|1)$ in the full symmetry algebra.
In particular the charges with the label $1$ act on the first entry in the
tensor product \eqref{bhtens}, while the charges with the label $2$ on the
second entry.

The full set of non-vanishing (anti-)commutation relations are then given
by\,\footnote{Here $\e_{12} = 1 = -\e_{21}$ is the usual antisymmetric
tensor.\fnsv}
\begin{align}
& [\mathfrak{J}_{_{U(1)}},\mathfrak{Q}_i] = \e_{ij} \mathfrak{Q}_i \ , \qquad \quad
[\mathfrak{J}_{_{U(1)}},\mathfrak{S}_i] = \e_{ij} \mathfrak{S}_i \ , \no
\\ & \{\mathfrak{Q}_i,\mathfrak{Q}_j\} = 2\d_{ij} \mathfrak{P} \ ,
\qquad \{\mathfrak{S}_i,\mathfrak{S}_j\} = 2\d_{ij} \mathfrak{K} \ ,
\qquad \{\mathfrak{Q}_i,\mathfrak{S}_j\} = 2\d_{ij} \mathfrak{C}\ ,
\end{align}
or alternatively in the complex basis
\begin{align}
& [\mathfrak{J}_{_{U(1)}},\mathfrak{G}_{\mathfrak{q},\mathfrak{s}\pm}] = \pm i\mathfrak{G}_{\mathfrak{q},\mathfrak{s}\pm} \ ,
\qquad \{\mathfrak{G}_{\mathfrak{q}_\pm}, \mathfrak{G}_{\mathfrak{q}\mp}\} = 2 \mathfrak{P} \ ,
\qquad \{\mathfrak{G}_{\mathfrak{s}_\pm}, \mathfrak{G}_{\mathfrak{s}\mp}\} = 2 \mathfrak{K} \ ,
\qquad \{\mathfrak{G}_{\mathfrak{q}_\pm}, \mathfrak{G}_{\mathfrak{s}\mp}\} = 2 \mathfrak{C} \ .
\end{align}

\refstepcounter{section}
\def\theequation{B.\arabic{equation}}
\setcounter{equation}{0}
\section*{Appendix B: Decomposition of the tensor product of two 2-dimensional representations}\label{appb}
\addcontentsline{toc}{section}{B \ Decomposition of the tensor product of two 2-dimensional representations}

In this appendix we give the explicit details of the decomposition of the
tensor product of two of the 2-dimensional representations of section \ref{massi}
following the construction in section \ref{32}. 
In this case we have
four states that are acted on as follows by the generators of the algebra:
\begin{align}
\nonumber
(\alg{C},\alg{P},\alg{K})\khh & = (C,P,K)\khh, &
(\alg{C},\alg{P},\alg{K})\khs & = (C,P,K)\khs,
\\\nonumber
\alg{Q}\khh & = a_1\ksh + \tilde a_2 \khs, &
\alg{Q}\khs & = a_1\kss + \tilde b_2 \khh,
\\\nonumber
\alg{S}\khh & = c_1\ksh + \tilde c_2 \khs, &
\alg{S}\khs & = c_1\kss + \tilde d_2 \khh,
\\ \nonumber \\
\nonumber
(\alg{C},\alg{P},\alg{K})\kss & = (C,P,K)\kss, &
(\alg{C},\alg{P},\alg{K})\ksh & = (C,P,K)\khs,
\\\nonumber
\alg{Q}\kss & = b_1\khs - \tilde b_2 \ksh, &
\alg{Q}\ksh & = b_1\khh - \tilde a_2 \kss,
\\
\alg{S}\kss & = d_1\khs - \tilde d_2 \ksh, &
\alg{S}\ksh & = d_1\khh - \tilde c_2 \kss,
\end{align}
where the labels $1$, $2$ refer to the first and second entry in the tensor
product and we recall that the action on the tensor product is given by the
coproduct \eqref{coproduct}, so that
\begin{equation}\begin{split}
& \tilde a_2 = a_2 U_1 \ , \qquad
\tilde b_2 = b_2 U_1 \ , \qquad
\tilde c_2 = c_2 U_1^{-1} \ , \qquad
\tilde d_2 = d_2 U_1^{-1} \ ,
\\ & 2C = 2C_1 + 2C_2 = a_1 d_1 + b_1 c_1 + \tilde a_2 \tilde d_2 + \tilde b_2 \tilde c_2 = a_1 d_1 + b_1 c_1 + a_2 d_2 + b_2 c_2 \ ,
\\ & P = P_1 + U_1^2 P_2 = a_1 b_1 + \tilde a_2 \tilde b_2 = a_1 b_1 + U_1^2 a_2 b_2 \ , \qquad
\\ & K = K_1 + U_1^{-2} K_2 = c_1 d_1 + \tilde c_2 \tilde d_2 = c_1 d_1 + U_1^{-2} c_2 d_2 \ .
\end{split}\end{equation}
These relations imply
\begin{equation}\begin{split}
M^2 = 4(C^2 - PK) & = (a_1 d_1 + b_1 c_2 + \tilde a_2 \tilde d_2 + \tilde b_2 \tilde c_2)^2 - 4(a_1 b_1 + \tilde a_2 \tilde b_2)(c_1 d_1 + \tilde c_2 \tilde d_2)
\\
& = (a_1 d_1 - b_1 c_1 + \tilde a_2 \tilde d_2 - \tilde b_2 \tilde c_2)^2 - 4(a_1 \tilde c_2 - c_1 \tilde a_2)(b_1 \tilde d_2 - d_1\tilde b_2) = M_b^2 - \ell_{ac} \ell_{bd}\ ,
\end{split}\end{equation}
where
\begin{equation}
M_b \equiv (a_1 d_1 - b_1 c_1 + \tilde a_2 \tilde d_2 - \tilde b_2 \tilde c_2) \ , \qquad
\ell_{ac} = 2(a_1 \tilde c_2 - c_1 \tilde a_2) \ , \qquad
\ell_{bd} = 2(b_1 \tilde d_2 - d_1 \tilde b_2) \ .
\end{equation}
It is then clear that the bound-state points occur when either $\ell_{ac} = 0$
or $\ell_{bd} = 0$. Furthermore, for the scattering of two physical states,
{\em i.e.} when the following reality conditions are satisfied
\begin{equation}
a_i^* = d_i \ , \qquad b_i^* = c_i \ , \qquad U_i^* = U_i^{-1} \ ,
\end{equation}
we find
\begin{equation}
C^* = C \ , \qquad P^* = K \ , \qquad M^* = M \ , \qquad
M_b^* = M_b^{\vphantom{*}} \ , \qquad \ell_{ac}^* = - \ell_{bd}^{\vphantom{*}} \ .
\end{equation}

To explicitly find the decomposition into two irreps, let us start by taking
\begin{equation}
|w_0\rangle = \khh \ , \qquad |\tilde w_0\rangle \equiv \frac 1M [\alg{Q},\alg{S}]|w_0\rangle =
\frac 1M \big[- M_b \khh + \ell_{ac} \kss\big] \ .
\end{equation}
It then follows that
\begin{equation}\label{sssss}
|\Phi_\pm\rangle = \frac1M\big[(M \mp M_b) \khh \pm \ell_{ac} \kss\big]\ .
\end{equation}
Alternatively we could have started by taking
\begin{equation}
|w_0\rangle = \kss \ ,
\end{equation}
in which case we end up with
\begin{equation}
|\Phi_\pm\rangle = \frac1M\big[(M \pm M_b) \kss \mp \ell_{bd} \khh\big]\ .
\end{equation}
It is easy to see that these states are proportional to each other from the identity
\begin{equation}\label{idmmb}
(M \pm M_b)(M\mp M_b) + \ell_{ac}\ell_{bd} =0\ .
\end{equation}
This same identity, along with the reality conditions, can be used to see that
\begin{equation}
\langle\Phi_\mp|\Phi_\pm\rangle = 0\ .
\end{equation}

Working with the state \eqref{sssss} we can apply the fermionic generators to find
\begin{equation}\begin{split}
\alg{Q}|\Phi_\pm\rangle
= & \frac1M \big[ \big( (M \mp M_b ) a_1 \mp \ell_{ac} \tilde b_2 \big) \khs
+ \big( (M \mp M_b ) \tilde a_2 \pm \ell_{ac} b_1 \big) \khs \big] \ ,\\
\alg{S}|\Phi_\pm\rangle
= & \frac1M \big[ \big( (M \mp M_b ) c_1 \mp \ell_{ac} \tilde d_2 \big) \khs
+ \big( (M \mp M_b ) \tilde c_2 \pm \ell_{ac} d_1 \big) \khs \big] \ .
\end{split}\end{equation}
One can then check that
\begin{equation}
\alg{Q}|\Phi_\pm\rangle \propto \alg{S}|\Phi_\pm\rangle \propto |\Psi_\pm\rangle \ .
\end{equation}
{\color{black} {\it \underline{Proof}}. This is seen explicitly from the following algebra:
\begin{equation*}\begin{split}
&\vphantom{\frac12}
((M\mp M_b)a_1 \mp \ell_{ac} \tilde b_2)
((M\mp M_b)\tilde c_2 \pm \ell_{ac} d_1)
-
((M\mp M_b)\tilde a_2 \pm \ell_{ac} b_1)
((M\mp M_b)c_1 \mp \ell_{ac} \tilde d_2)
\\ = &\vphantom{\frac12}
(M \mp M_b)^2(a_1 \tilde c_2 - \tilde a_2 c_1)
\pm (M\pm M_b)\ell_{ac}(a_1 d_1 - b_1 c_1 - \tilde a_2 \tilde d_2 - \tilde b_2 \tilde c_2)
+ \ell_{ac}^2(b_1 \tilde d_2 - d_1 \tilde b_2)
\\ = &
\frac12 (M \mp M_b)^2\ell_{ac} \pm (M\pm M_b)\ell_{ac} M_b + \frac12\ell_{ac}^2\ell_{bd}
\\ = &
\frac12 (M^2 + M_b^2)\ell_{ac}\mp M M_b \ell_{ac} + M_b^2 \ell_{ac} \pm M M_b\ell_{ac} + \frac12(M_b^2 - M^2)\ell_{ac} = 0 \ .
\end{split}\end{equation*}\fnsv}

Furthermore,
\begin{equation}\begin{split}
\alg{Q}\alg{S}|\Phi_\pm\rangle
= & \frac1{2M}(2C \pm M) \big[(M\mp M_b)\khh \pm\ell_{ac}\kss\big]\ ,
\\
\alg{S}\alg{Q}|\Phi_\pm\rangle
= & \frac1{2M}(2C\mp M) \big[(M \mp M_b)\khh \pm \ell_{ac}\kss\big]\ .
\end{split}\end{equation}
{\color{black}  {\it \underline{Proof}}. The explicit derivation is

{\footnotesize \begin{equation*}\begin{split}
\alg{Q}\alg{S}|\Phi_\pm\rangle
= & \frac1M \big[
\big((M \mp M_b)(b_1 c_1 + \tilde b_2 \tilde c_2) \mp \ell_{ac} (b_1 \tilde d_2 - d_1 \tilde b_2)\big)\khh
+ \big((M \mp M_b)(a_1 \tilde c_2 - c_1 \tilde a_2) \pm \ell_{ac}(a_1 d_1 + \tilde a_2 \tilde d_2) \big)\kss\big]
\\
= & \frac1{2M} \big[
\big((M \mp M_b)(2C-M_b) \mp (M_b^2 - M^2)\big)\khh + (M \pm c)\ell_{ac}\kss\big]\ ,
\\
= & \frac1{2M}(2C \pm M) \big[
(M\mp M_b)\khh \pm\ell_{ac}\kss\big]\ ,
\\
\alg{S}\alg{Q}|\Phi_\pm\rangle
= & \frac1M \big[
\big((M \mp M_b)(a_1 d_1 + \tilde a_2 \tilde d_2) \pm \ell_{ac} (b_1 \tilde d_2 - d_1 \tilde b_2)\big)\khh
+ \big(-(M \mp M_b)(a_1 \tilde c_2 - c_1 \tilde a_2) \pm \ell_{ac}(b_1 c_1 + \tilde b_2 \tilde c_2) \big)\kss\big]
\\
= & \frac1{2M} \big[
\big((M \mp M_b)(2C+M_b) \pm (M_b^2 - M^2)\big)\khh - (M \mp c)\ell_{ac}\kss\big]\ ,
\\
= & \frac1{2M}(2C\mp M) \big[(M \mp M_b)\khh \pm \ell_{ac}\kss\big]\ .
\end{split}\end{equation*}\fnsv}}

\corp{Then using} 
\eqref{idmmb} it is clear that $\alg{Q}\alg{S}|\Phi_\pm\rangle
\propto |\Phi_\pm\rangle$ and $\alg{S}\alg{Q}|\Phi_\pm\rangle \propto |\Phi_\pm
\rangle$ and hence this shows explicitly that
$\{|\Phi_\pm\rangle,|\Psi_\pm\rangle\}$ form two 2-dimensional irreps.

At the bound-state points $\ell_{ac} =0$ or $\ell_{bd}=0$ one of the
irreducible blocks contains $|\phi\phi\rangle$, as either $|\Phi_+\rangle$ or
$|\Phi_-\rangle$ aligns to this state. Therefore, one can focus on the $\phi \,
\phi \rightarrow \phi \, \phi$ entry of the S-matrix (supplemented by the
appropriate dressing phase) to ascertain whether this corresponds to a pole in
the s-channel in the physical region.




\begin{thebibliography}{30}

\parskip=0.pt

\bibitem{rev}
N.~Beisert, C.~Ahn, L.~F.~Alday, Z.~Bajnok, J.~M.~Drummond, L.~Freyhult, N.~Gromov and R.~A.~Janik {\it et al.},
``Review of AdS/CFT Integrability: An Overview,''
Lett.\ Math.\ Phys.\ {\bf 99} (2012) 3
[arXiv:1012.3982].

{\color{black}
\bibitem{rev1}
G.~Arutyunov and S.~Frolov,
``Foundations of the $AdS_5 \times S^5$ Superstring. Part I,''
J.\ Phys.\ A {\bf 42}, 254003 (2009)
[arXiv:0901.4937].
}
\bibitem{sym}
K.~Zarembo,
``Strings on Semisymmetric Superspaces,''
JHEP {\bf 1005} (2010) 002
[arXiv:1003.0465].
\\
L.~Wulff,
``Superisometries and integrability of superstrings,''
JHEP {\bf 1405} (2014) 115
[arXiv:1402.3122].

\bibitem{ads2}
I.~R.~Klebanov and A.~A.~Tseytlin,
``Intersecting M-branes as four-dimensional black holes,''
Nucl.\ Phys.\ B {\bf 475} (1996) 179
[arXiv:hep-th/9604166].
\\
A.~A.~Tseytlin,
``Harmonic superpositions of M-branes,''
Nucl.\ Phys.\ B {\bf 475} (1996) 149
[arXiv:hep-th/9604035].
\\
M.~J.~Duff, H.~Lu and C.~N.~Pope,
``$AdS_5 \times S^5$ untwisted,''
Nucl.\ Phys.\ B {\bf 532} (1998) 181
[arXiv:hep-th/9803061].
\\
H.~J.~Boonstra, B.~Peeters and K.~Skenderis,
``Brane intersections, anti-de Sitter space-times and dual superconformal theories,''
Nucl.\ Phys.\ B {\bf 533} (1998) 127
[arXiv:hep-th/9803231].
\\
J.~Lee and S.~Lee,
``Mass spectrum of D=11 supergravity on $AdS_2 \times S^2 \times T^7$,''
Nucl.\ Phys.\ B {\bf 563} (1999) 125
[arXiv:hep-th/9906105].

\bibitem{adscft}
J.~M.~Maldacena,
``The Large $N$ limit of superconformal field theories and supergravity,''
Adv.\ Theor.\ Math.\ Phys.\ {\bf 2} (1998) 231
[arXiv:hep-th/9711200].
\\
E.~Witten,
``Anti-de Sitter space and holography,''
Adv.\ Theor.\ Math.\ Phys.\ {\bf 2} (1998) 253
[arXiv:hep-th/9802150].

\bibitem{dual}
A.~Strominger,
``$AdS_2$ quantum gravity and string theory,''
JHEP {\bf 9901} (1999) 007
[arXiv:hep-th/9809027].
\\
G.~W.~Gibbons and P.~K.~Townsend,
``Black holes and Calogero models,''
Phys.\ Lett.\ B {\bf 454} (1999) 187
[arXiv:hep-th/9812034].
\\
J.~M.~Maldacena, J.~Michelson and A.~Strominger,
``Anti-de Sitter fragmentation,''
JHEP {\bf 9902} (1999) 011
[arXiv:hep-th/9812073].
\\
C.~Chamon, R.~Jackiw, S.-Y.~Pi and L.~Santos,
``Conformal quantum mechanics as the CFT$_1$ dual to AdS$_2$,''
Phys.\ Lett.\ B {\bf 701} (2011) 503
[arXiv:1106.0726].

\bibitem{Metsaev:1998it}
R.~R.~Metsaev and A.~A.~Tseytlin,
``Type IIB superstring action in $AdS_5 \times S^5$ background,''
Nucl.\ Phys.\ B {\bf 533} (1998) 109
[arXiv:hep-th/9805028].

\bibitem{sc}
J.-G.~Zhou,
``Super 0-brane and GS superstring actions on $AdS_2 \times S^2$,''
Nucl.\ Phys.\ B {\bf 559} (1999) 92
[arXiv:hep-th/9906013].
\\
N.~Berkovits, M.~Bershadsky, T.~Hauer, S.~Zhukov and B.~Zwiebach,
``Superstring theory on $AdS_2 \times S^2$ as a coset supermanifold,''
Nucl.\ Phys.\ B {\bf 567} (2000) 61
[arXiv:hep-th/9907200].

\bibitem{Bena:2003wd}
I.~Bena, J.~Polchinski and R.~Roiban,
``Hidden symmetries of the $AdS_5 \times S^5$ superstring,''
Phys.\ Rev.\ D {\bf 69} (2004) 046002
[arXiv:hep-th/0305115].

\bibitem{Green:1984sg}
M.~B.~Green and J.~H.~Schwarz,
``Anomaly Cancellation in Supersymmetric D=10 Gauge Theory and Superstring Theory,''
Phys.\ Lett.\ B {\bf 149} (1984) 117.

\bibitem{Sorokin:2011rr}
D.~Sorokin, A.~Tseytlin, L.~Wulff and K.~Zarembo,
``Superstrings in $AdS_2 \times S^2 \times T^6$,''
J.\ Phys.\ A {\bf 44} (2011) 275401
[arXiv:1104.1793].

\bibitem{Cagnazzo:2011at}
A.~Cagnazzo, D.~Sorokin and L.~Wulff,
``More on integrable structures of superstrings in $AdS_4 \times \mathbb{C}\mathbf{P}^3$ and $AdS_2 \times S^2 \times T^6$ superbackgrounds,''
JHEP {\bf 1201} (2012) 004
[arXiv:1111.4197].

\bibitem{lcg}
C.~G.~Callan, Jr., H.~K.~Lee, T.~McLoughlin, J.~H.~Schwarz, I.~Swanson and X.~Wu,
``Quantizing string theory in $AdS_5 \times S^5$: Beyond the pp wave,''
Nucl.\ Phys.\ B {\bf 673} (2003) 3
[hep-th/0307032].
\\
S.~Frolov, J.~Plefka and M.~Zamaklar,
``The $AdS_5 \times S^5$ superstring in light-cone gauge and its Bethe equations,''
J.\ Phys.\ A {\bf 39} (2006) 13037
[arXiv:hep-th/0603008].
\\
G.~Arutyunov, S.~Frolov and M.~Zamaklar,
``Finite-size Effects from Giant Magnons,''
Nucl.\ Phys.\ B {\bf 778} (2007) 1
[arXiv:hep-th/0606126].

\bibitem{Berenstein:2002jq}
D.~E.~Berenstein, J.~M.~Maldacena and H.~S.~Nastase,
``Strings in flat space and pp waves from $\mathcal{N}=4$ Super Yang Mills,''
JHEP {\bf 0204} (2002) 013
[arXiv:hep-th/0202021].

\bibitem{Murugan:2012mf}
J.~Murugan, P.~Sundin and L.~Wulff,
``Classical and quantum integrability in AdS$_{2}$/CFT$_{1}$,''
JHEP {\bf 1301} (2013) 047
[arXiv:1209.6062].

\bibitem{amsw}
M.~C.~Abbott, J.~Murugan, P.~Sundin and L.~Wulff,
``Scattering in AdS$_2$/CFT$_1$ and the BES Phase,''
JHEP {\bf 1310} (2013) 066
[arXiv:1308.1370].

\bibitem{beis0}
N.~Beisert,
``The $SU(2|2)$ dynamic S-matrix,''
Adv.\ Theor.\ Math.\ Phys.\ {\bf 12} (2008) 945
[arXiv:hep-th/0511082].

\bibitem{conv}
G.~Arutyunov, S.~Frolov and M.~Zamaklar,
``The Zamolodchikov-Faddeev algebra for $AdS_5 \times S^5$ superstring,''
JHEP {\bf 0704} (2007) 002
[arXiv:hep-th/0612229].

\bibitem{an}
C.~Ahn and R.~I.~Nepomechie,
``Review of AdS/CFT Integrability, Chapter III.2: Exact World-Sheet S-Matrix,''
Lett.\ Math.\ Phys.\ {\bf 99}, 209 (2012)
[arXiv:1012.3991].

\bibitem{Borsato:2013qpa}
R.~Borsato, O.~Ohlsson Sax, A.~Sfondrini, B.~Stefanski and A.~Torrielli,
``The all-loop integrable spin-chain for strings on AdS$_3 \times S^3 \times T^4$: the massive sector,''
JHEP {\bf 1308} (2013) 043
[arXiv:1303.5995].

\bibitem{HT}
B.~Hoare and A.~A.~Tseytlin,
``Massive S-matrix of $AdS_3 \times S^3 \times T^4$ superstring theory with mixed 3-form flux,''
Nucl.\ Phys.\ B {\bf 873} (2013) 395
[arXiv:1304.4099].

\bibitem{Sfond}
A.~Sfondrini,
``Towards integrability for AdS$_3$/CFT$_2$,''
[arXiv:1406.2971].

\bibitem{Janik:2006dc}
R.~A.~Janik,
``The $AdS_5 \times S^5$ superstring worldsheet S-matrix and crossing symmetry,''
Phys.\ Rev.\ D {\bf 73} (2006) 086006
[arXiv:hep-th/0603038].

\bibitem{ads3phase}
R.~Borsato, O.~Ohlsson Sax, A.~Sfondrini, B.~Stefanski and A.~Torrielli,
``Dressing phases of AdS$_3$/CFT$_2$,''
Phys.\ Rev.\ D {\bf 88} (2013) 066004
[arXiv:1306.2512].
\\
A.~Babichenko, A.~Dekel and O.~Ohlsson Sax,
``Finite-gap equations for strings on $AdS_3 \times S^3 \times T^4$ with mixed 3-form flux,''
[arXiv:1405.6087].

\bibitem{hl}
R.~Hernandez and E.~Lopez,
``Quantum corrections to the string Bethe ansatz,''
JHEP {\bf 0607} (2006) 004
[arXiv:hep-th/0603204].

\bibitem{bes}
N.~Beisert, B.~Eden and M.~Staudacher,
``Transcendentality and Crossing,''
J.\ Stat.\ Mech.\ {\bf 0701} (2007) P01021
[arXiv:hep-th/0610251].

\bibitem{pert}
T.~Klose, T.~McLoughlin, R.~Roiban and K.~Zarembo,
``Worldsheet scattering in $AdS_5 \times S^5$,''
JHEP {\bf 0703} (2007) 094
[arXiv:hep-th/0611169].
\\
P.~Sundin and L.~Wulff,
``Worldsheet scattering in AdS$_3$/CFT$_2$,''
JHEP {\bf 1307} (2013) 007
[arXiv:1302.5349].
\\
B.~Hoare and A.~A.~Tseytlin,
``On string theory on $AdS_3 \times S^3 \times T^4$ with mixed 3-form flux: tree-level S-matrix,''
Nucl.\ Phys.\ B {\bf 873} (2013) 682
[arXiv:1303.1037].

\bibitem{Arutyunov:2009pw}
G.~Arutyunov, M.~de Leeuw and A.~Torrielli,
``On Yangian and Long Representations of the Centrally Extended $\mathfrak{su}(2|2)$ Superalgebra,''
JHEP {\bf 1006} (2010) 033
[arXiv:0912.0209].

\bibitem{pr}
M.~Grigoriev and A.~A.~Tseytlin,
``Pohlmeyer reduction of $AdS_5 \times S^5$ superstring sigma model,''
Nucl.\ Phys.\ B {\bf 800} (2008) 450
[arXiv:0711.0155].
\\
B.~Hoare and A.~A.~Tseytlin,
``Towards the quantum S-matrix of the Pohlmeyer reduced version of $AdS_5 \times S^5$ superstring theory,''
Nucl.\ Phys.\ B {\bf 851}, 161 (2011)
[arXiv:1104.2423].
\\
R.~Shankar and E.~Witten,
``The S Matrix of the Supersymmetric Nonlinear Sigma Model,''
Phys.\ Rev.\ D {\bf 17} (1978) 2134.
\\
K.-i.~Kobayashi and T.~Uematsu,
``S-matrix of $\mathcal{N}=2$ supersymmetric Sine-Gordon theory,''
Phys.\ Lett.\ B {\bf 275} (1992) 361
[hep-th/9110040].

\bibitem{BogdanLatest}
R.~Borsato, O.~Ohlsson Sax, A.~Sfondrini and B.~Stefanski,
``All-loop worldsheet S matrix for $AdS_3 \times S^3 \times T^4$,''
[arXiv:1403.4543].
\\
R.~Borsato, O.~Ohlsson Sax, A.~Sfondrini and B.~Stefanski,
``The complete $AdS_3 \times S^3 \times T^4$ worldsheet S-matrix,''
[arXiv:1406.0453].

\bibitem{Arutyunov:2006ak}
G.~Arutyunov, S.~Frolov, J.~Plefka and M.~Zamaklar,
``The Off-shell Symmetry Algebra of the Light-cone $AdS_5 \times S^5$ Superstring,''
J.\ Phys.\ A {\bf 40} (2007) 3583
[hep-th/0609157].

\bibitem{GotzQuellaSchomerus}
G.~Gotz, T.~Quella and V.~Schomerus,
``Representation theory of $\mathfrak{sl}(2|1)$,''
J.\ Algebra {\bf 312} (2007) 829
[arXiv:hep-th/0504234].

\bibitem{IoharaKoga}
K.~Iohara and Y.~Koga,
``Central extensions of Lie superalgebras,''
Commentarii Mathematici Helvetici {\bf 76} (2001) 110.

\bibitem{tor}
C.~Gomez and R.~Hernandez,
``The magnon kinematics of the AdS/CFT correspondence,''
JHEP {\bf 0611} (2006) 021
[arXiv:hep-th/0608029].
\\
J.~Plefka, F.~Spill and A.~Torrielli,
``On the Hopf algebra structure of the AdS/CFT S-matrix,''
Phys.\ Rev.\ D {\bf 74} (2006) 066008
[arXiv:hep-th/0608038].

\bibitem{beis1}
N.~Beisert,
``An $SU(1|1)$-invariant S-matrix with dynamic representations,''
Bulg.\ J.\ Phys.\ {\bf 33S1}, 371 (2006)
[arXiv:hep-th/0511013].
\\
N.~Beisert,
``The Analytic Bethe Ansatz for a Chain with Centrally Extended $\mathfrak{su}(2|2)$ Symmetry,''
J.\ Stat.\ Mech.\ {\bf 0701} (2007) P01017
[arXiv:nlin/0610017].

\bibitem{Zamolodchikov:1994za}
A.~B.~Zamolodchikov,
``Thermodynamics of imaginary coupled sine-Gordon: Dense polymer finite size scaling function,''
Phys.\ Lett.\ B {\bf 335} (1994) 436.

\bibitem{Arutyunov:2007tc}
  G.~Arutyunov and S.~Frolov,
  ``On String S-matrix, Bound States and TBA,''
  JHEP {\bf 0712} (2007) 024
  [arXiv:0710.1568].
  
\bibitem{Arutyunov:2009kf}
  G.~Arutyunov and S.~Frolov,
  ``The Dressing Factor and Crossing Equations,''
  J.\ Phys.\ A {\bf 42} (2009) 425401
  [arXiv:0904.4575].


\bibitem{uc}
L.~Bianchi, V.~Forini and B.~Hoare,
``Two-dimensional S-matrices from unitarity cuts,''
JHEP {\bf 1307} (2013) 088
[arXiv:1304.1798].
\\
O.~T.~Engelund, R.~W.~McKeown and R.~Roiban,
``Generalized unitarity and the worldsheet S matrix in $AdS_n \times S^n \times M^{10-2n}$,''
JHEP {\bf 1308} (2013) 023
[arXiv:1304.4281].
\\
L.~Bianchi and B.~Hoare,
``$AdS_3 \times S^3 \times M^4$ string S-matrices from unitarity cuts,''
JHEP {\bf 1408} (2014) 097
[arXiv:1405.7947].

\cortwo{
\bibitem{dr}
  B.~Hoare, A.~Stepanchuk and A.~A.~Tseytlin,
  ``Giant magnon solution and dispersion relation in string theory in $AdS_3 \times S^3 \times T^4$ with mixed flux,''
  Nucl.\ Phys.\ B {\bf 879} (2014) 318
  [arXiv:1311.1794].
\\
  T.~Lloyd, O.~O.~Sax, A.~Sfondrini and B.~Stefanski,
  ``The complete worldsheet S matrix of superstrings on $AdS_3 \times S^3 \times T^4$ with mixed three-form flux,''
  [arXiv:1410.0866].}

\bibitem{Drinfeld:1987sy}
V.~G.~Drinfeld,
``A new realization of Yangians and quantized affine algebras,''
Sov.\ Math.\ Dokl.\ {\bf 36} (1988) 212.

\bibitem{Spill:2008tp}
F.~Spill and A.~Torrielli,
``On Drinfeld's second realization of the AdS/CFT $\mathfrak{su}(2|2)$ Yangian,''
J.\ Geom.\ Phys.\ {\bf 59} (2009) 489
[arXiv:0803.3194].

\bibitem{Beisert:2007ds}
N.~Beisert,
``The S-matrix of AdS/CFT and Yangian symmetry,''
PoS SOLVAY {\bf } (2006) 002
[arXiv:0704.0400].

\bibitem{Torrielli:2011gg}
A.~Torrielli,
``Yangians, S-matrices and AdS/CFT,''
J.\ Phys.\ A {\bf 44} (2011) 263001
[arXiv:1104.2474].

\bibitem{Zamol2}
A.~B.~Zamolodchikov and A.~B.~Zamolodchikov,
``Massless factorized scattering and sigma models with topological terms,''
Nucl.\ Phys.\ B {\bf 379} (1992) 602.

\bibitem{Matsumoto:2007rh}
T.~Matsumoto, S.~Moriyama and A.~Torrielli,
``A Secret Symmetry of the AdS/CFT S-matrix,''
JHEP {\bf 0709} (2007) 099
[arXiv:0708.1285].

\bibitem{Beisert:2007ty}
N.~Beisert and F.~Spill,
``The Classical r-matrix of AdS/CFT and its Lie Bialgebra Structure,''
Commun.\ Math.\ Phys.\  {\bf 285} (2009) 537
[arXiv:0708.1762].

\bibitem{Beisert:2014hya}
N.~Beisert and M.~de Leeuw,
``The RTT-Realization for the Deformed $\alg{gl}(2|2)$ Yangian,''
[arXiv:1401.7691].
  
\bibitem{Pittelli:2014ria}
A.~Pittelli, A.~Torrielli and M.~Wolf,
``Secret Symmetries of Type IIB Superstring Theory on $AdS_3 \times S_3 \times M_4$,''
[arXiv:1406.2840].

\end{thebibliography}
\end{document}